\documentclass[reprint,amsmath,amssymb,aps,showpacs,superscriptaddress,twocolumn,longbibliography,prl]{revtex4-1}
\usepackage{graphicx}
\usepackage{bm}
\usepackage{epsfig}
\usepackage{amssymb}
\usepackage{amsfonts}
\usepackage{braket}
\usepackage{color}
\usepackage{epstopdf}
\usepackage{hyperref}
\usepackage{float}
\restylefloat{table}
\usepackage{bibentry}
\usepackage{multirow}
\usepackage[caption=false]{subfig}

\graphicspath{{figures/}}

\newcommand{\ba}{\begin{eqnarray}}
\newcommand{\ea}{\end{eqnarray}}
\newcommand{\bd}{\begin{displaymath}}
\renewcommand{\v}[1]{{\bf #1}}
\newcommand{\nn}{\nonumber \\}

\begin{document}
\title{ Gapless Kitaev Spin Liquid to Classical String Gas through Tensor Networks }

\author{Hyun-Yong Lee}
\affiliation{Institute for Solid State Physics, University of Tokyo, Kashiwa, Chiba 277-8581, Japan}

\author{Ryui Kaneko}
\affiliation{Institute for Solid State Physics, University of Tokyo, Kashiwa, Chiba 277-8581, Japan}

\author{Tsuyoshi Okubo}
\affiliation{Department of Physics, University of Tokyo, Tokyo 113-0033, Japan}

\author{Naoki Kawashima}
\email{kawashima@issp.u-tokyo.ac.jp}
\affiliation{Institute for Solid State Physics, University of Tokyo, Kashiwa, Chiba 277-8581, Japan}
\date{\today}

\begin{abstract}
	We provide a framework for understanding the gapless Kitaev spin liquid\,(KSL) in the language of tensor network\,(TN). Without introducing Majorana fermion, most of the features of the KSL including the symmetries, gauge structure, criticality and vortex-freeness are explained in a compact TN representation. Our construction reveals a hidden string gas structure of the KSL. With only two variational parameters to adjust, we obtain an accurate KSL ansatz with the bond dimension $D=8$ in a compact form, where the energy is about $0.007\%$ higher than the exact one. 
	\end{abstract}
\maketitle

{\it Introduction-} Quantum spin liquids\,(QSL) represent a state of quantum matter which is not characterized by any local order parameters even at zero temperature. These novel states are expected to exhibit long-range entanglement leading to the topological order and fractionalized excitations\cite{Lucile2017}. For example, the nearest-neighbor resonating valence bond\,(nnRVB) states\cite{Anderson87} have been extensively studied as variational wavefunctions for the ground states of frustrated quantum magnets\cite{Moessner2001,Wen2002}. Indeed, the nnRVB states are topologically ordered\cite{Moessner2001,Wen2002,Zhou2016a,Poiblanc2012, HY2017} and support spinon excitations carrying the fractionalized quantum number\cite{Wen2002,Zhou2016a}. However, since they are not exact ground states of the antiferromagnetic Heisenberg model, variational methods with the nnRVB states have been employed to search for true ground states\cite{Kaneko2014,Jiang2016,Iqbal16,Mei17}. The Haldane phase, which is also known as a symmetry-protected topological phase, is another fascinating phase one can find in the $S=1$ quantum spin chain. The novel character that discriminates the Haldane phase from trivial gapped states was most clearly revealed by the discovery of Affleck–Kennedy–Lieb–Tasaki\,(AKLT) model and its exact ground state or AKLT state\cite{AKLT1987}. 
The compact representation of AKLT state\cite{Klumper1993} provided a new insight into the Haldane phase. In addition, it was subsequently used in a variety of contexts for variational calculations on the quantum spin systems\cite{Klumper1993, Kolezhuk1998}.

Kitaev honeycomb model\,(KHM) is an exactly soluble model which exhibits gapless and gapped KSL ground states with fractionalized excitations\cite{Kitaev2006}. Recent successful realizations of Kitaev materials\cite{Khaliullin2005,Jackeli2009, Plumb2014, Zhou2016b, Banerjee2016, Trebst2017, Singh2010} triggered a burst of theoretical investigations on KHM and its extensions\cite{Kimchi11,Catuneanu2018,Gohlke2018}. In addition, due to the non-Abelian phase of the KSL driven by the magnetic field\cite{Jiang11,Zhu18} and its potential application to quantum computation, it has been a subject of active research for the last decade. We refer the readers to Ref.\cite{Nayak2008} for an exhaustive list of relevant literature. The TN methods have been also employed to represent the KSL\cite{Schmoll2017,Osorio2014}. However, the Majorana basis TN requires a three-dimensional structure which makes it impractical as a tool for the numerical optimization\cite{Schmoll2017}. On the other hand, the spin basis TN study, which was done with the computationally expensive optimization, suffers from undesirable breaking of symmetries\cite{Osorio2014}. In this Letter, we provide a compact TN representation for KHM that is defined with the spin basis and retains various symmetries.

\begin{figure}[!b]
  \includegraphics[width=0.4\textwidth]{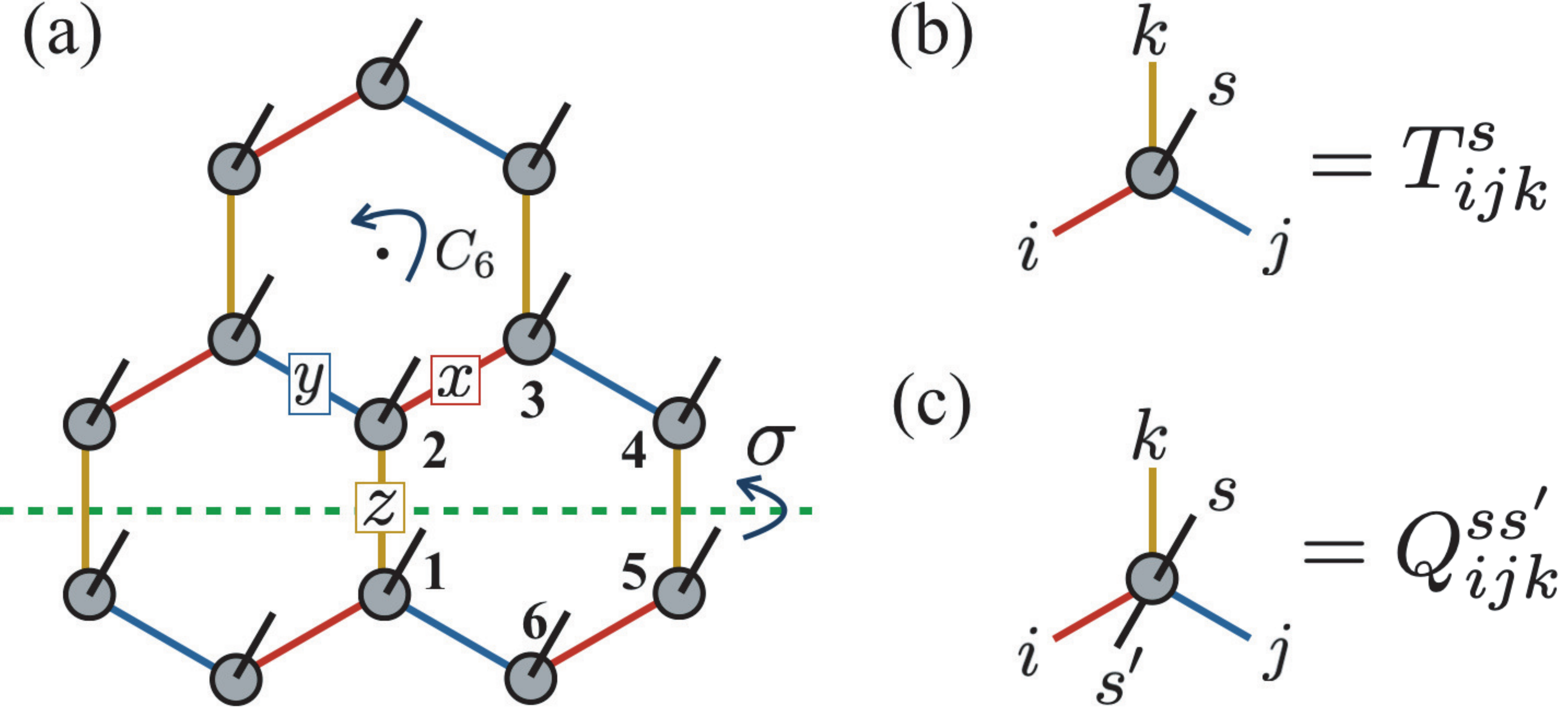}
  \caption{ Schematic figures of (a) TPS setup on the honeycomb lattice, (b) its building block tensor $T_{ijk}^s$, and (c) a building block tensor $Q_{ijk}^{ss'}$ of the loop gas TPO defined in Eq.\,\eqref{eq:q_operator}. Here, the $x$-, $y$- and $z$-links in the model\,[Eq.\,\eqref{eq:hamiltonian}] are characterized by red, blue and yellow colors, respectively. }
  \label{fig:schematic}
\end{figure}

{\it Model.} The KHM is defined as\cite{Kitaev2006}
\begin{align}
	\hat{\mathcal{H}} = - \sum_{\langle \alpha\beta \rangle_\gamma } J_{\gamma} \hat{\sigma}_\alpha^\gamma \hat{\sigma}_\beta^\gamma
	\label{eq:hamiltonian}
\end{align} 
where $\langle \alpha\beta \rangle_{\gamma}$ stands for a pair on the $\gamma\,(=x,y,z)$ links connecting sites $\alpha$ and $\beta$ as depicted in Fig.\,\ref{fig:schematic}\,(a). As demonstrated in Kitaev's seminal work\cite{Kitaev2006}, the Hamiltonian commutes with the so-called flux operators defined on all hexagonal plaquette\,($p$): $[\hat{\mathcal{H}},\hat{W}_p]=0$ with $\hat{W}_p = \hat{\sigma}^x_1 \hat{\sigma}^y_2 \hat{\sigma}^z_3 \hat{\sigma}^x_4 \hat{\sigma}^y_5 \hat{\sigma}^z_6$ where the site indices 1-6 are defined in Fig.\,\ref{fig:schematic}\,(a). Therefore, the Hilbert space is sectorized by each flux number $\{ \hat{W}_p = \pm 1 \}$. Even further, in each sector, the KHM becomes a noninteracting Majorana fermion hopping model in the background of static $Z_2$ gauge fields. The ground states live in the vortex-free sector\,($\hat{W}_p = +1$ for all $p$), which form a critical KSL phase around the isotropic point\,($J_x=J_y=J_z=\pm1$). In this Letter, we only consider the isotropic point at which the KHM is invariant under the following symmetry transformations: $C_6 \hat{U}_{C_6}$ and $\sigma \hat{U}_{\sigma}$, where $ \hat{U}_{C_6}  = \bigotimes_\alpha \left( \hat{\sigma}^0_\alpha + i\hat{\sigma}^x_\alpha + i\hat{\sigma}^y_\alpha + i\hat{\sigma}^z_\alpha \right)/2$, $\hat{U}_{\sigma}  =  \bigotimes_\alpha \left( \hat{\sigma}^x_\alpha - \hat{\sigma}^y_\alpha \right)/\sqrt{2}$, and $C_6$, $\sigma$ respectively denote the $60^\circ$ spatial rotation and inversion as depicted in Fig.\,\ref{fig:schematic}\,(a). 
One can easily verify $[ C_6 \hat{U}_{C_6}, \hat{\mathcal{H}}] = 0 = [ \sigma \hat{U}_{\sigma}, \hat{\mathcal{H}}]$. Also, the KHM is invariant under the time-reversal and translational symmetries.


{\it Tensor network representation.} We employ the tensor product state\,(TPS) representation\,\cite{Frank2008}. Since the KSL is a zero-flux state\,\cite{You2008}, we reasonably assume the TPS to be translationally symmetric. Additionally, we assume that the tensor does not depend on the sublattice, and therefore our ansatz is rewritten as 
%
	$|\psi\rangle = {\rm tTr} \prod_{\alpha} | T_{i_{\alpha} j_{\alpha} k_{\alpha}} \rangle,	$
%
where tTr stands for the tensor trace or contraction of all virtual indices $\{i_\alpha, j_\alpha, k_\alpha\}$, $\alpha$ labels the site, $|T_{ijk}\rangle \equiv \sum_s T_{ijk}^s|s\rangle$ with $s$ being the local quantum number. Its graphical illustration is presented in Fig.\,\ref{fig:schematic}, where the black open leg denotes the physical degrees of freedom. In what follows, we will construct the local tensor $|T_{ijk}\rangle$\,(we identify $T_{ijk}^s$ and $|T_{ijk}\rangle$ and call both ``tensor" hereafter) with consideration for the symmetries, the vortex-free condition and gauge structure. 
In the main text, we only discuss the ferromagnetic model\,$(J_\gamma=1)$, since the antiferromagnetic one is a trivial generalization and discussed in the supplementary material\,(SM)\,\cite{SM}.

{\it Zeroth order tensor.} We begin with introducing a bond dimension $D=2$ tensor product operator\,(TPO) referred to as the {\it loop gas}\,(LG) operator, $ \hat{Q}_{\rm LG} = {\rm tTr} \prod_{\alpha} Q_{i_{\alpha} j_{\alpha} k_{\alpha}}^{ss'}|s\rangle\langle s'|$ with a building block tensor
\begin{eqnarray}
	Q_{i j k}^{s s'} = \tau_{i j k } [(\hat{\sigma}^x)^{1-i} (\hat{\sigma}^y)^{1-j} (\hat{\sigma}^z)^{1-k}]_{ss'},
	\label{eq:q_operator}
\end{eqnarray}
which is depicted in Fig.\,\ref{fig:schematic}\,(c). The virtual indices $i,j$ and $k$ range from 0 to 1\,($D=2$), and non-zero elements of $\tau$-tensor are $\tau_{000} = -i$ and $\tau_{011}=\tau_{101}=\tau_{110}=1$. To simplify the notation, we define a local operator $\hat{Q}_{ijk} = \sum_{s,s'} Q_{ijk}^{ss'}|s\rangle\langle s'|$. 
One can verify in the local tensor level that the LG operator respects the symmetries of KHM. For instance, applying $C_6 \hat{U}_{C_6}$ on the $\hat{Q}$-tensor leaves it intactly, i.e.,
\begin{align}
	(C_6 \hat{U}_{C_6}) \hat{Q}_{ijk} (C_6 \hat{U}_{C_6})^{\dagger} = \hat{Q}_{ijk}.\nonumber
\end{align}
Here, we use the facts that the $\hat{U}_{C_6}$-transformation rotates the spin, i.e., $U_{C_6} \hat{\sigma}^{x,y,z}_\alpha U_{C_6}^{\dagger} = \hat{\sigma}^{z,x,y}_\alpha $, while the $C_6$-rotation permutes the virtual indices as followse $C_6 \circ (ijk) = (kij)$. Therefore, the resulting LG operator is invariant under the $(C_6 \hat{U}_{C_6})$-transformation, and other symmetries of KHM can be shown in a similar way\cite{SM}. Note that the $Q$-operator satisfies the following relation 
\begin{eqnarray}
	\hat{\sigma}^x \hat{Q}_{ijk} &=& v_{jj'} v_{kk'}^* \hat{Q}_{ij'k'},\nonumber\\
	\hat{\sigma}^y \hat{Q}_{ijk} &=& v_{kk'} v_{ii'}^* \hat{Q}_{i'jk'},\nonumber\\
	\hat{\sigma}^z \hat{Q}_{ijk} &=& v_{ii'} v_{jj'}^* \hat{Q}_{ij'k'},
	\label{eq:d2_xlink}
\end{eqnarray}
with a matrix $v$, of which non-zero elements are $v_{01} = i$ and $v_{10}=1$, acting on the virtual bonds. Repeated indices are summed over, except where explicitly stated otherwise. Using Eq.\,\eqref{eq:d2_xlink}, one can verify a relation $\hat{W}_p \hat{Q}_{\rm LG} = \hat{Q}_{\rm LG} \hat{W}_p = \hat{Q}_{\rm LG}$. To be more specific, the invariance of a patch of $\hat{Q}_{\rm LG}$ under the action of $\hat{W}_p$ can be shown as follows: 
\begin{align}
	\includegraphics[width=0.45\textwidth]{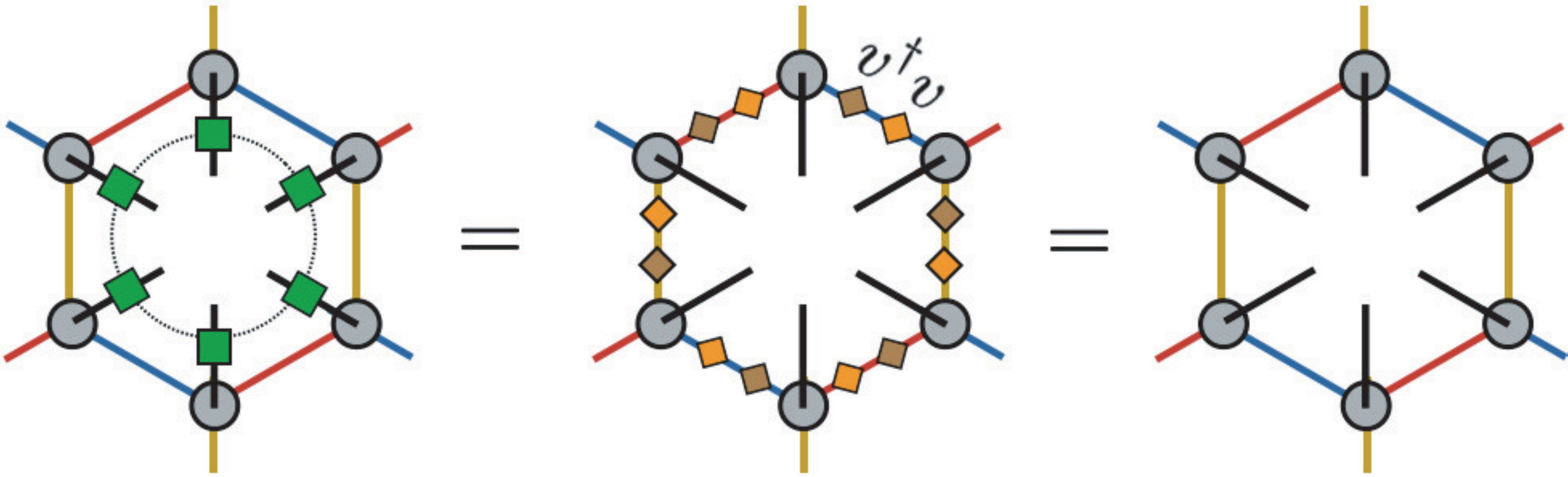},\nonumber
\end{align}
where the connected green squares denote $\hat{W}_p$, and a physical leg is omitted for simplicity.
Here, Eq.\,\eqref{eq:d2_xlink} and $v^{\dagger}v = 1$ are used in the first and second equalities, respectively. This remarkable relation guarantees a quantum state $|\psi\rangle = \hat{Q}_{\rm LG} |\phi\rangle$, where $|\phi\rangle$ is an arbitrary state, being vortex-free and thus non-magnetic. Notice that the LG operator is identical to the projector $\prod_p (1+\hat{W}_p)/2$ up to a normalization factor.

Regarding the $(C_6 \hat{U}_{C_6})$-symmetry, let us apply $\hat{Q}_{\rm LG}$ on a product state $|\phi_0\rangle = \otimes_\alpha |(111)\rangle_\alpha $, where $|(111)\rangle$ denotes a spin aligned along (1,1,1) direction: $\langle(111)|\,\vec{\sigma}\,|(111)\rangle = (1,1,1)/\sqrt{3}$. Note that the ansatz $|\phi_0\rangle$ is a classical ground state respecting the $(C_6 \hat{U}_{C_6})$-symmetry. Now, we define a quantum state $|\psi_0\rangle \equiv \hat{Q}_{\rm LG}|\phi_0\rangle$ which consists of a building block tensor
\begin{align}
	|T^0_{ijk}\rangle \equiv \hat{Q}_{ijk} |(111)\rangle.
	\label{eq:parent_tensor}
\end{align}
We refer to it as {\it zeroth order tensor}. By virtue of the $\tau$-tensor in $\hat{Q}_{ijk}$, one can visualize the ansatz $|\psi_0\rangle$ as follows
\begin{align}
	\includegraphics[width=0.48\textwidth]{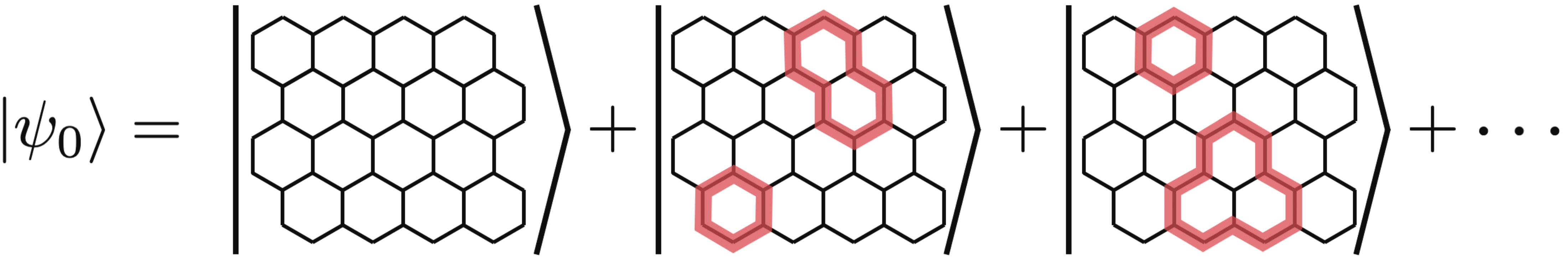}
	\label{eq:string_net}.
\end{align}
Here, the empty site stands for $|(111)\rangle$ state while the red loops denote the product of $\hat{\sigma}^x|(111)\rangle, \hat{\sigma}^y|(111)\rangle$ and $\hat{\sigma}^z|(111)\rangle$ states depending on the direction of loop on each site.

By computing the norm of the LG ansatz, we can show its criticality. To this end, we first note that the LG operator is hermitian as well as idempotent\cite{SM}: $\hat{Q}_{\rm LG}^\dagger \hat{Q}_{\rm LG} = N_{\Gamma} \hat{Q}_{\rm LG}$ where $N_\Gamma$ is a total number of the loop configuration in the system. Using such properties and a simple identity $\langle (111)| \hat{\sigma}^\gamma |(111)\rangle = 1/\sqrt{3}$, it is straightforward to show that the norm of $|\psi_0\rangle$ reads 
\begin{align}
	\langle \psi_0 | \psi_0\rangle = N_{\Gamma} \sum_{G\in \Gamma} \left(\frac{1}{\sqrt{3}}\right)^{l_G}
	= N_{\Gamma} \times Z_{O(1)}\left(\frac{1}{\sqrt{3}}\right),
\end{align}
where $\Gamma$ denotes a set of all possible loop configurations and $l_G$ is a total length of loops in a configuration $G$. Also, $Z_{O(1)}(x)$ stands for the partition function of the classical $O(1)$ loop gas model with the fugacity $x$, which is exactly solvable and critical at $x_c = 1/\sqrt{3}$\,\cite{Nienhuis1982}. It indicates that the norm of $|\psi_0\rangle$ is exactly mapped into the partition function of the critical classical model which guarantees the criticality of $|\psi_0\rangle$\cite{Ardonne2004}. In addition, the Ising conformal field theory\,(CFT) with the central charge $c=1/2$ is known to characterize the critical LG model\cite{Nienhuis1982}, which is consistent with the KSL of KHM\cite{Teresia13, Lahtinen14, Meichanetzidis16}.

The LG structure encoded in the $\tau$-tensor is useful in describing the vortex excitation of the KSL. To see this, we first note that the $\tau$-tensor is invariant under a gauge transformation $g=\hat{\sigma}^z$, i.e. $g_{ii'}g_{jj'} g_{kk'}\tau_{i'j'k'} = \tau_{ijk}$, and thus
\begin{align}
	g_{ii'} g_{jj'} g_{kk'}|T^0_{i'j'k'}\rangle = |T^0_{ijk}\rangle. \nonumber
\end{align}
With a trivial gauge transformation $I_2$ being a two-dimensional identity matrix, they form a $Z_2$ invariant gauge group\,(IGG). String-like action of $g$ on links would twist the gauge fields\cite{Kitaev2006} along the string and hence create two vortices $\hat{W}_p=-1$ at both ends as demonstrated below:
\begin{align}
	\includegraphics[width=0.33\textwidth]{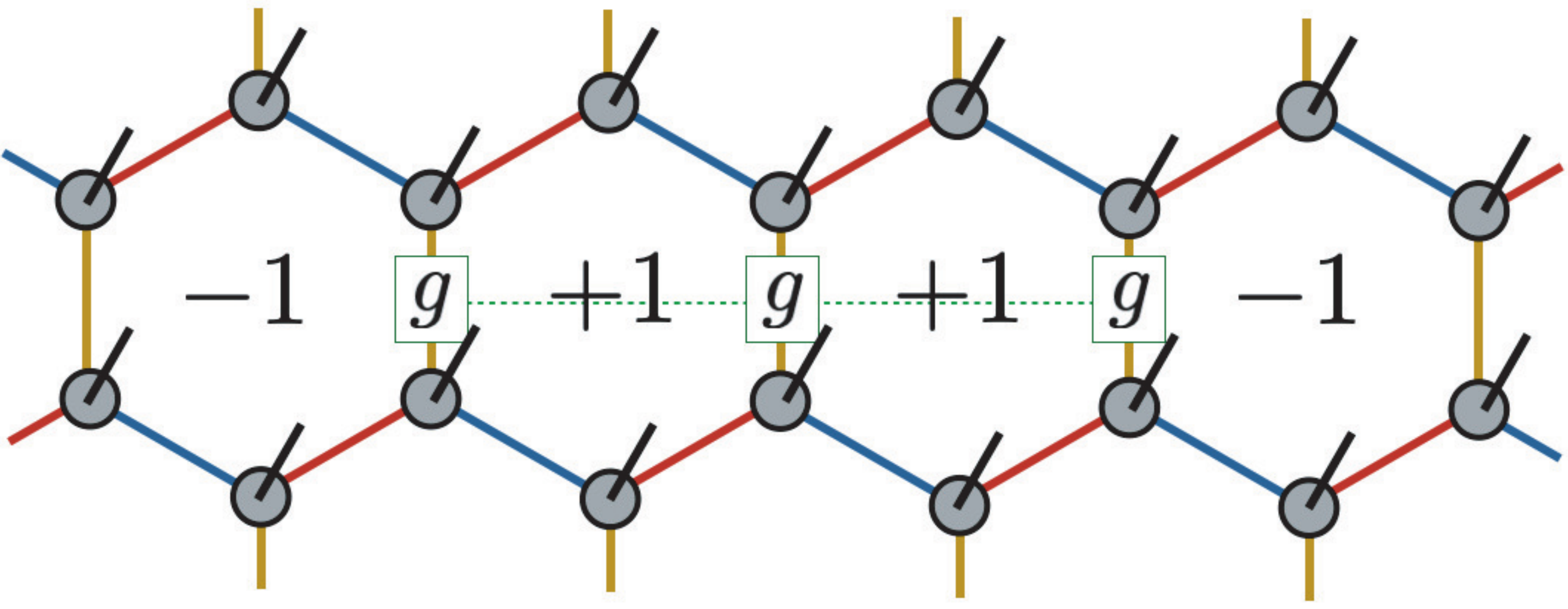},\nonumber
\end{align}
where $\pm 1$ in the hexagon denotes $\hat{W}_p$. One can explicitly show\cite{SM} such creation and move of fluxes using Eq.\,\eqref{eq:d2_xlink}. 

Finally, we measure the KHM energy\,(per bond) of $|\psi_0\rangle$ and obtain $E=-0.16349$ which is rather higher than the exact one $E_{\rm Kitaev} \simeq -0.19682$\cite{Kitaev2006}. Details in numerics will be discussed later. By construction, the LG ansatz $|\psi_0\rangle$ made of zeroth order tensor satisfies most of the physical constraints respected in the KSL\,[see SM for the time-reversal and $\sigma \hat{U}_{\sigma}$ symmetries] but is energetically far away from the exact solution. In what follows, we present a simple but effective TPO\,($D=2$) applied to the LG ansatz which reduces the energy greatly without violating the constraints. We refer to it as the {\it dimer gas}\,(DG) operator.

{\it Higher order tensors.} The DG operator is defined by $\hat{R}_{\rm DG} = {\rm tTr} \prod_{\alpha} \hat{R}_{i_\alpha j_\alpha k_\alpha} $ with
\begin{align}
	\hat{R}_{ijk} = \zeta_{i j k}(\hat{\sigma}^x)^i (\hat{\sigma}^y)^j (\hat{\sigma}^z)^k.
	\label{eq:r_operator}
\end{align}
Here, non-zero elements of $\zeta$-tensor are $\zeta_{000}=1$ and $\zeta_{100}=\zeta_{010}=\zeta_{001}=z$ with $i,j,k=0,1$, and $z$ is a real(or pure imaginary) variational parameter fixing the fugacity of a dimer. In this context, the dimer denotes the operator $\hat{\mathcal{H}}_{\alpha\beta}^\gamma$. Then, the DG operator can be interpreted as a sum of all possible dimer configurations, i.e., $\hat{R}_{\rm DG}(z) = \sum_{G\in \Gamma_{\rm D}} \hat{R}_G(z) $ where $\hat{R}_G(z) =\bigotimes_{\langle\alpha\beta\rangle_\gamma \in G} (z \hat{\mathcal{H}}_{\alpha\beta}^\gamma)$ is defined for each dimer configuration $G$, and $\Gamma_{\rm D}$ is the set of all dimer configurations:
\begin{align}
	\includegraphics[width=0.48\textwidth]{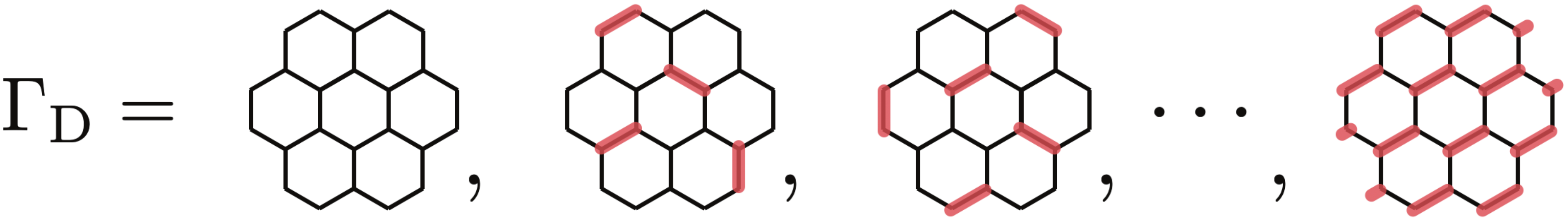}
	\label{eq:p_operator}.
\end{align}
Due to $[\hat{\mathcal{H}}_{\alpha\beta}^\gamma, \hat{W}_p] = 0$, it is obvious that $\hat{R}_G$ commutes with $\hat{W}_p$ for any $G$, and hence $\hat{R}_{\rm DG}$ does; $[\hat{R}_{\rm DG},\hat{W}_p]=0$. In fact, we can easily prove $[\hat R_{\rm DG}, \hat{Q}_{\rm LG}]=0$ and that the DG operator respects all symmetries of the KSL\,\cite{SM}. Therefore, its multiplication to $|\psi_0\rangle$ does not contaminate the features of the KSL regardless of $z$. 
Moreover, it can be expressed as the polynomial function of the KHM Hamiltonian, which may be the reason why it improves the energy of the ansatz quite efficiently. The first key observation is that we can graphically represent Eq.\,\eqref{eq:hamiltonian} raised to the $n$-th power as the linear combination of elements of $\Gamma_{\rm D}$ as
\begin{align}
	\includegraphics[width=0.48\textwidth]{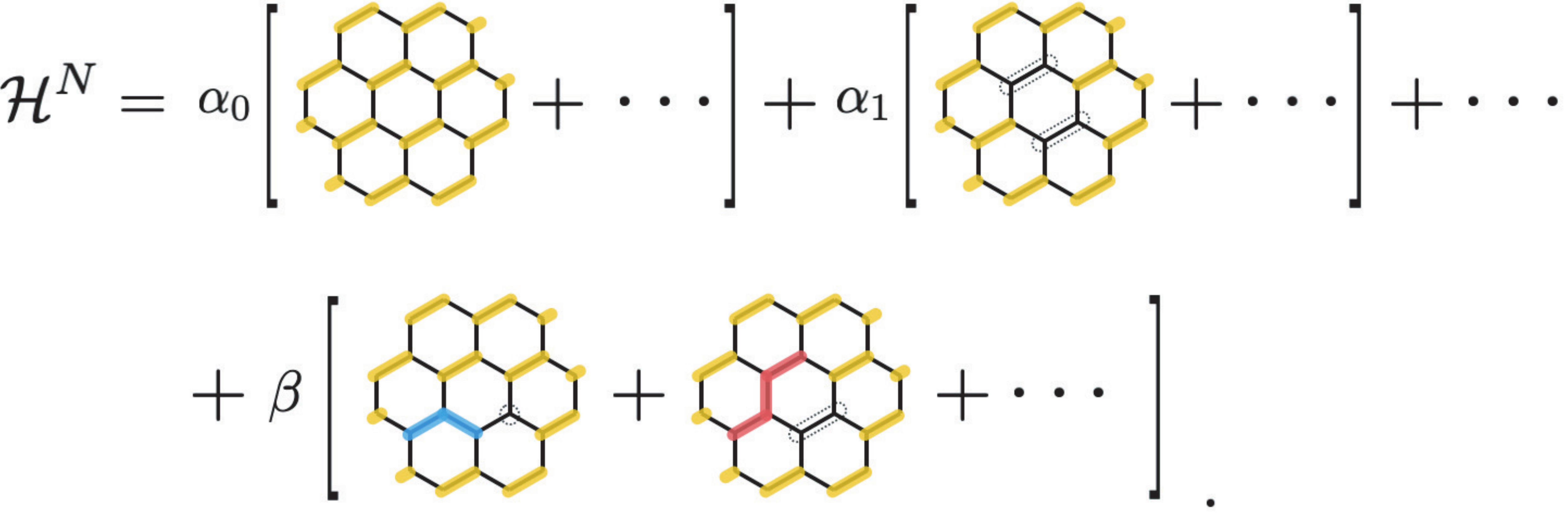}\nonumber
\end{align}
Here, the number of sites in the system is assumed to be $2N$. The terms grouped with a coefficient $\alpha_0$ are the fully-packed configurations while the second ones are configuration with $N-2$ dimers. The terms on the second line have $q$-mers longer than dimer, e.g. trimer $\hat{\mathcal{H}}_{\alpha\beta}^x \hat{\mathcal{H}}_{\beta \gamma}^y$. All those terms with $q$-mers are canceled by the anticommutativity of Pauli matrices, and thus $\beta = 0$. Note that the configurations with the same number of dimers share the coefficient which resembles the $\hat{R}_{\rm DG}$. Then, one can recast it as $\hat{R}_{\rm DG} = \sum_M h_M \mathcal{\hat{H}}^{N-2M} $ with proper coefficients $h_M$. Note that our approach is not a perturbative one\cite{Vanderstraeten17}. 

\begin{figure}[!b]
	\includegraphics[width=0.5\textwidth]{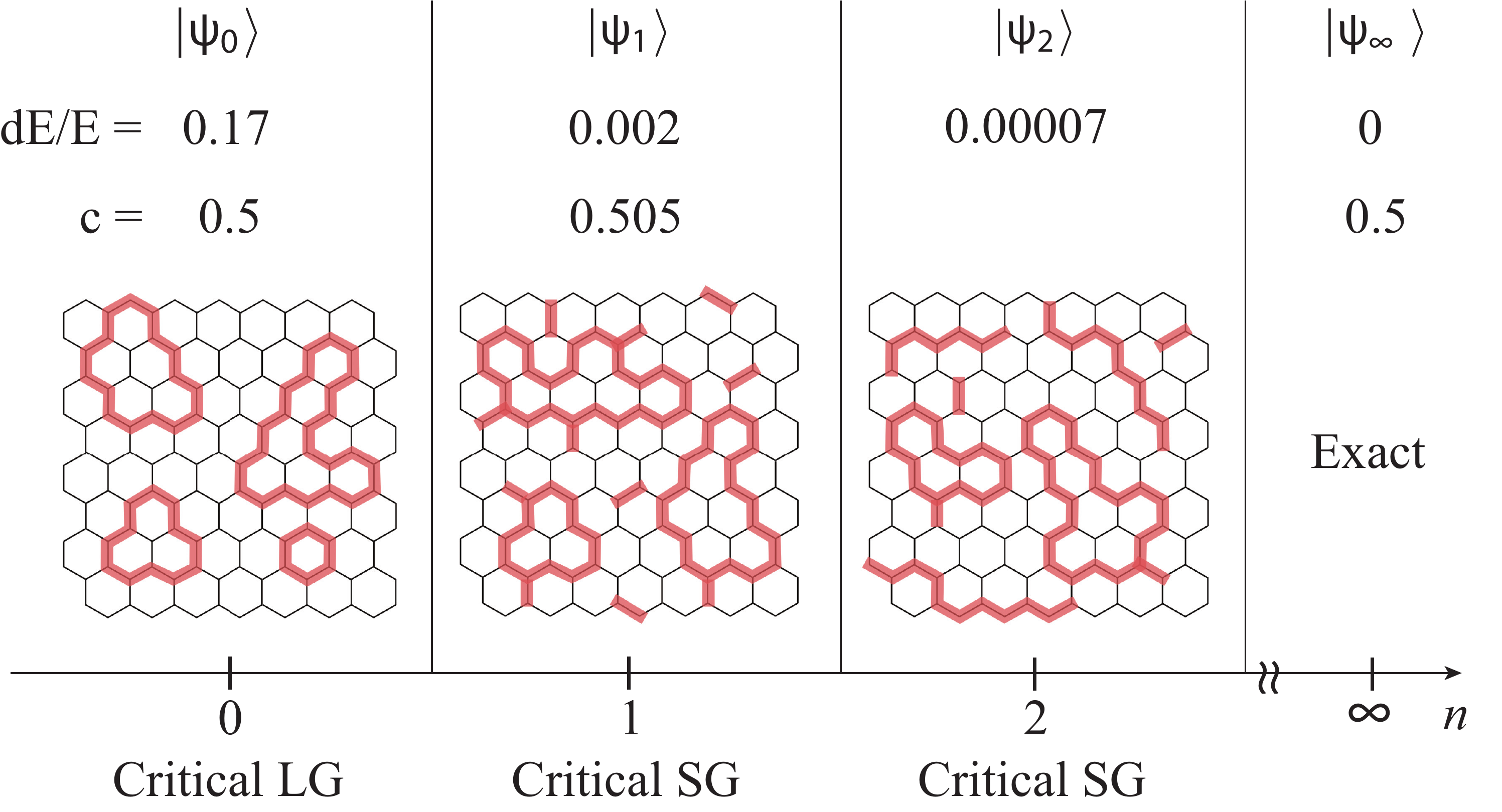}
	\caption{Overview of the $n$-th order ansatz $|\psi_n\rangle$ obtained by LG and DG operators, where SG denotes the string gas, $dE = E-E_{\rm Kitaev}$ the energy deviation, and $c$ stands for the central charge.  }
	\label{fig:string_config}
\end{figure}

Now, we define the $n$-th order ansatz as $|\psi_n(\{z_i\})\rangle = \prod_{i=1}^n\hat{R}_{\rm DG}(z_i) |\psi_0\rangle$ having $n$ complex variational parameters. Due to the application of the DG operator, the ansatz $|\psi_n\rangle$ can be interpreted as a {\it string gas} state which is a linear superposition of {\it string} configurations. The string configuration consists of open and closed strings, connected loops and string-connected loops as depicted in Fig.\,\ref{fig:string_config}. The building block tensor of $|\psi_n\rangle$, referred to as the $n$-th order tensor, is obtained by applying the $\hat{R}_{ijk}$-operator $n$-times on the zeroth order tensor in Eq.\,\eqref{eq:parent_tensor}. The bond dimension scales as $D=2^{n+1}$.
Note that the LG feature or $\tau$-tensor in the zeroth tensor is inherited by all higher order tensors. Furthermore, the $\hat{R}_{ijk}$-operator is invariant only under the trivial gauge transformation, and thus its action does not enlarge the $Z_2$ IGG of which the non-trivial element is simply
%
	$g_n = I_{2^n} \otimes \hat{\sigma}^z$.	
%
In contrast to the zeroth order case, the norm of $|\psi_n\rangle$ does not map to the LG model. However, by employing the loop TN renormalization\cite{Shuo17}, we numerically prove that the $n$-th order ansatz are also critical and characterized by the Ising CFT as summarized in Fig.\,\ref{fig:string_config}. We also present the best variational energies at each order in Fig.\,\ref{fig:string_config}, and the details are given below.


{\it Variational ansatz.} Now, we turn on and tune variational parameters to obtain a better ansatz than the zeroth one. We parametrize the $\zeta$-tensor as follows: $\zeta_{000} = \cos\phi$, $\zeta_{100} = \zeta_{010} = \zeta_{001} = \sin\phi$, and hence $\hat{R}_{ijk}(z) \rightarrow \hat{R}_{ijk}(\phi)$. For measuring the energy, we employ the corner transfer matrix renormalization group method\,(CTMRG)\cite{Nishino1996,Orus2009,Corboz2010} of which accuracy is controlled by the dimension $\chi$ of CTM. The parallel C++ library mptensor\cite{mptensor} is utilized to perform CTMRG.


%
\begin{figure}[!t]
  \includegraphics[width=0.5\textwidth]{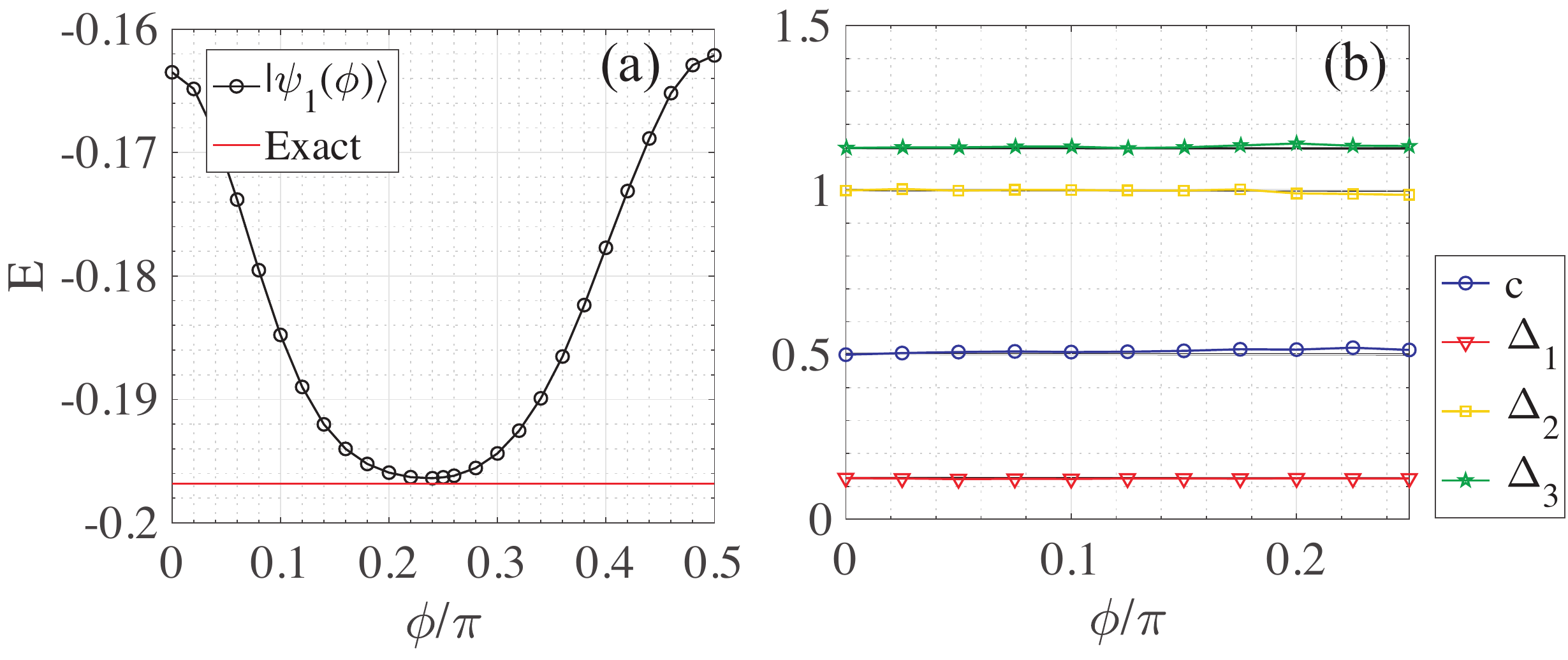}
  \caption{ (a) Energy of $|\psi_1(\phi)\rangle$ of which the building block tensor is defined in Eq.\,\eqref{eq:d4_tensor}, and (b) the central charge $c$ and scaling dimensions $\Delta_i$ as a function of $\phi$. Black solid lines in (b) denote the exact ones from the Ising universality class.}
  \label{fig:d4_results}
\end{figure}

Let us begin with the first order ansatz $|\psi_1(\phi)\rangle$ and its building block tensor
\begin{align}
	|T_{i_1 j_1 k_1 }^1(\phi)\rangle = \hat{R}_{i j k}(\phi) |T_{i_0 j_0 k_0}^0\rangle,
	\label{eq:d4_tensor}
\end{align}
where $i_n = 2^n i + i_{n-1}$ and $i = 0,1$. The energy of $|\psi_1(\phi)\rangle$ is presented in Fig.\,\ref{fig:d4_results}\,(a) as a function of $\phi$, of which the lowest $E = -0.19644$ is found at $\phi = 0.24\pi$. Here, we fix $\chi=64$. It is remarkable that the first order tensor\,($D=4$) already attains such a small error of $0.2\%$. Furthermore, we perform the loop TN renormalization to evaluate the norm of $| \psi_1(\phi)\rangle$ and extract the central charge and scaling dimensions presented in Fig.\,\ref{fig:d4_results}\,(b)[see SM for more details]. All those are in excellent agreement with the ones of Ising CFT, and therefore our ansatze are critical and belong to the same universality class. 
%
\begin{figure}[!t]
  \includegraphics[width=0.5\textwidth]{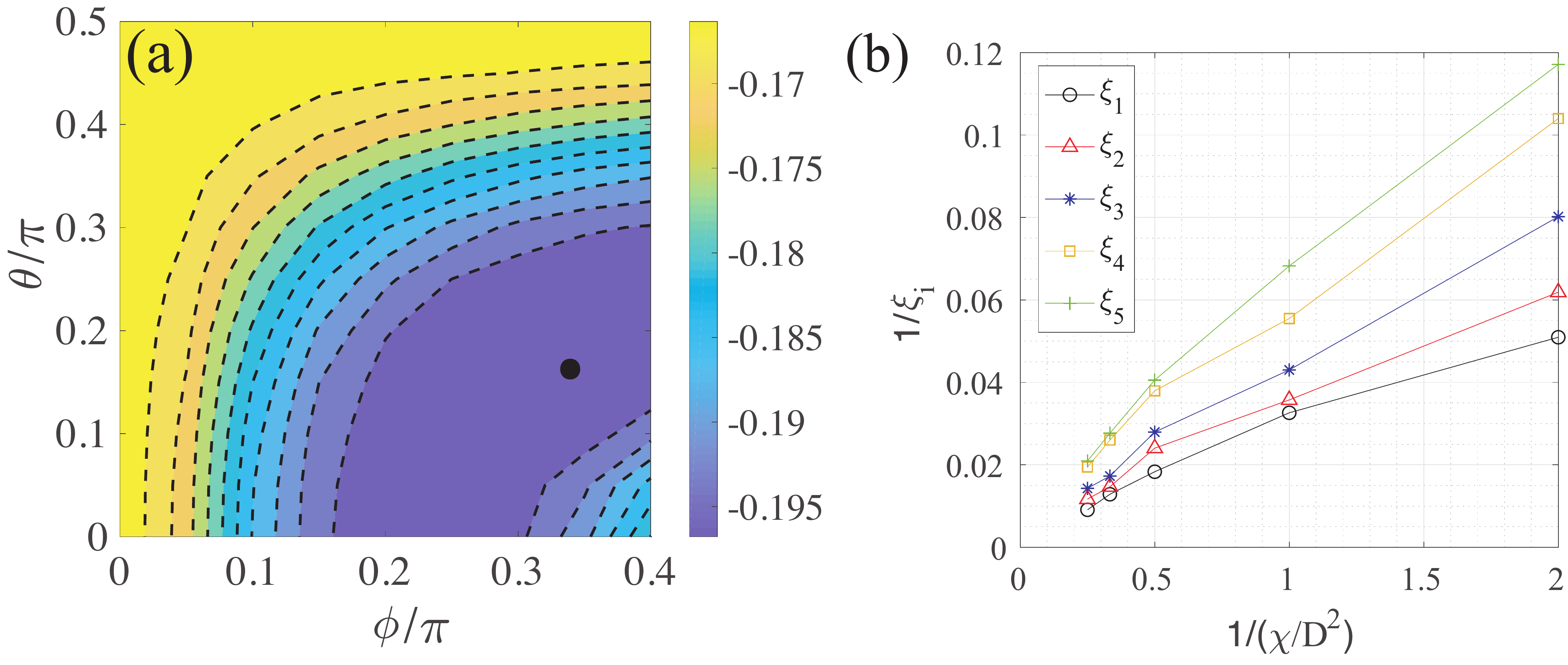}
  \caption{ (a) Energy landscape of $|\psi_2(\phi,\theta)\rangle$ constructed by the tensor in Eq.\,\eqref{eq:d8_tensor} as functions of $\phi$ and $\theta$. The energy minima is denoted by the black dot, at which the variational energy is $E=-0.19681$\cite{SM}. (b) The five largest correlation lengths\,$\xi_i$ of the best ansatz represented by the dot in (a) where $\chi$ stands for the bond dimension of CTMRG.  }
  \label{fig:d8_energy}
\end{figure}
To obtain an ansatz even closer to the KSL, we consider the second order ansatz $|\psi_2\rangle$ and tensor\,($D=8$):
\begin{align}
	|T_{i_2 j_2 k_2 }^2(\phi,\theta)\rangle = \hat{R}_{i j k}(\theta) |T_{i_1 j_1 k_1}^1(\phi)\rangle.	
	\label{eq:d8_tensor}
\end{align}
Its overall energy landscape is shown in Fig.\,\ref{fig:d8_energy}\,(a) as functions of $(\phi,\theta)$ and minimized at $(\phi,\theta) = (0.342\pi,0.176\pi)$. After an additional scaling with respect to $\chi$\cite{SM}, we obtain the best variational ansatz with $E=-0.19681$ which is only $0.007\%$ higher than the exact one. Also, using the environment tensors\cite{hylee18}, the five largest correlation lengths\,($\xi_i$) are extracted and shown in Fig.\,\ref{fig:d8_energy}\,(b), which are diverging with $\chi$. Analogous figure is shown for $\psi_1$ in SM. Therefore, we reasonably conclude that the ansatz made of higher order tensors form a family of gapless states which we believe are smoothly connected to the exact KSL and, as a series, converge to it. 

Further, we found\cite{SM} that applying the (111)-direction magnetic field drives the ansatze into the gapped phase\cite{Kitaev2006}. We speculate that these gapped ansatze host non-Abelian anyonic excitations. The description of the non-Abelian and Abelian topological phases under the LG and SG schemes is an interesting question, and now further study is in progress\cite{hylee19}.

{\it Conclusion.} Based on the physical and gauge symmetries and the vortex-free condition, we have constructed the compact TN representation, which generates a family of KSL-like states sharing the features of the KHM ground state. In this sense, the ansatze given in this study are analogous to the AKLT state as a member of the Haldane states or the RVB state as an ansatz of frustrated quantum magnets\cite{AKLT1987,Wen2002}. Under this scheme, the string gas structure of the KSL comes in sight clearly which offers a novel viewpoint for the KSL and its physics. It also provides an intuitive picture for the KSL in the {\it spin} language without referring to the Majorana fermion, which has never been provided before. There are many generalizations that one can envision as well as concrete open questions involving the LG and SG ansatze, e.g., general LGs having larger internal degrees of freedom and their parent Hamiltonians. The relation between the general LGs and the string-net states\cite{Fendley2008} is another interesting question to ask. We also find that the ansatz discussed in the present Letter provides a good initial state for variational method for the  KHM with the magnetic field\cite{Ryui19}. Further, for the anisotropic KHM, one can choose the initial magnetic state which differs from the state $|(111)\rangle$ and introduce a bond-dependent dimer fugacity as additional variational parameters to optimize the model\cite{hylee19}. Therefore, we expect our work could furnish a better understanding of KSL and its neighboring phases observed in the Kitaev quantum magnets such as $\alpha$-RuCl$_3$\,\cite{Banerjee2016,Banerjee2018} and studied theoretically in extended KHMs\cite{Jackeli2009, Chaloupka2010, Singh2010, Kimchi11, Gohlke2018}. Using two variational parameters, the accuracy of $0.007\%$ in energy is obtained, which has never been achieved by other numerical optimizations\cite{Phien2015, Corboz2016, Frank2016}. This high accuracy, together with the observed systematic convergence, leads us to believe that the present scheme not only correctly captures the essence of KSL physics, but also provides a new direction for quantitatively accurate description of quantum spin liquids.

\acknowledgements
{\it Acknowledgements-} The computation in the present work was executed on computers at the Supercomputer Center, ISSP, University of Tokyo, and also on K-computer (project-ID: hp180225). N.K.'s work is funded by ImPACT Program of Council for Science, Technology and Innovation (Cabinet Office, Government of Japan).	H.-Y.L. was supported by MEXT as ``Exploratory Challenge on Post-K computer"\,(Frontiers of Basic Science: Challenging the Limits). R.K. and T.O. were supported by MEXT as ``Priority Issue on Post-K computer" (Creation of New Functional Devices and High-Performance Materials to Support Next-Generation Industries), and JSPS KAKENHI No. 15K17701, respectively.

\bibliography{reference.bib}

\clearpage
\onecolumngrid
\begin{center}
\textbf{\large  Supplemental Material: Gapless Kitaev Spin Liquid to Classical String Gas through Tensor Networks}
\end{center}

\setcounter{equation}{0}
\setcounter{figure}{0}
\setcounter{table}{0}
\setcounter{page}{1}

\begin{center}
\parbox[t][4cm][s]{0.8\textwidth}{		In this supplemental material, we prove in details that the loop gas\,(LG) and dimer gas\,(DG) operators proposed in the main text are invariant under the $(C_6 \hat{U}_{C_c})$-symmetry, $(\sigma \hat{U}_\sigma)$-symmetry and time-reversal symmetry transformations. Furthermore, we show that the DG operator can be recast as the polynomial function of the Kitaev Hamiltonian. Also, we explicitly show that inserting the non-trivial element of the $Z_2$ invariant gauge group between two tensors creates the vortex. Next, we compute the norm of the zeroth order ansatz analytically and discuss the conformal data of the norm of the higher order ansatze obtained by the loop tensor network renormalization. The variational energy with the complex dimer fugacity is discussed, and we give a simple generalization of our scheme to the antiferromagnetic Kitaev model. Finally, we present how the ansatz in the presence of the $(111)$-direction magnetic field is obtained.}
\end{center}

\section{ Symmetries of the Loop Gas Operator }

In the main text, we define the loop gas\,(LG) operator $ \hat{Q}_{\rm LG} = {\rm tTr} \prod_{\alpha} \hat{Q}_{i_{\alpha} j_{\alpha} k_{\alpha}}$ with the building block tensor
\begin{align}
	\hat{Q}_{i j k} = \tau_{i j k } (\hat{\sigma}^x)^{1-i} (\hat{\sigma}^y)^{1-j} (\hat{\sigma}^z)^{1-k},
\end{align}
where $i,j,k=0,1$, and the non-zero elements of $\tau$-tensor are
\begin{align}
	\tau_{000} = -i,\quad \tau_{011} = \tau_{101} = \tau_{110} = 1.
\end{align}
We consider unitary operators
\begin{align}
	\hat{U}_{C_6}  =  \frac{1}{2}\left( \hat{\sigma}^0 + i\hat{\sigma}^x + i\hat{\sigma}^y + i\hat{\sigma}^z \right),\quad\quad
	\hat{U}_{\sigma} = \frac{i}{\sqrt{2}} \left( \hat{\sigma}^x - \hat{\sigma}^y \right),
\end{align}
which transform the Pauli matrices in the following way:

\begin{align}
	& \hat{U}_{C_6} \hat{\sigma}^x	 \hat{U}_{C_6}^{\dagger} = \hat{\sigma}^z,\quad
	\hat{U}_{C_6} \hat{\sigma}^y \hat{U}_{C_6}^{\dagger} = \hat{\sigma}^x,\quad
	\hat{U}_{C_6} \hat{\sigma}^z \hat{U}_{C_6}^{\dagger} = \hat{\sigma}^y,\nn
	& \hat{U}_{\sigma} \hat{\sigma}^x \hat{U}_{\sigma}^{\dagger} = -\hat{\sigma}^y,\quad
	\hat{U}_{\sigma} \hat{\sigma}^y \hat{U}_{\sigma}^{\dagger} = -\hat{\sigma}^x,\quad
	\hat{U}_{\sigma} \hat{\sigma}^z \hat{U}_{\sigma}^{\dagger} = -\hat{\sigma}^z.
	\label{eq:unitary_rotations}
\end{align}
Let us see how the LG operator transforms locally under the $\hat{U}_{C_6}$ and $\hat{U}_{\sigma}$ rotations:

\begin{align}
	&\hat{U}_{C_6} \hat{Q}_{ijk} \hat{U}_{C_6}^{\dagger} 
	= \tau_{i j k } \hat{U}_{C_6}  (\hat{\sigma}^x)^{1-i} (\hat{\sigma}^y)^{1-j} (\hat{\sigma}^z)^{1-k} \hat{U}_{C_6}^{\dagger}
	= \tau_{i j k } (\hat{\sigma}^z)^{1-i} (\hat{\sigma}^x)^{1-j} (\hat{\sigma}^y)^{1-k},\nonumber\\
	&\hat{U}_{\sigma} \hat{Q}_{ijk} \hat{U}_{\sigma}^{\dagger} 
	= \tau_{i j k } \hat{U}_{\sigma} (\hat{\sigma}^z)^{1-i} (\hat{\sigma}^x)^{1-j} (\hat{\sigma}^y)^{1-k} \hat{U}_{\sigma}^{\dagger}
	= \tau_{i j k } (-\hat{\sigma}^y)^{1-i} (-\hat{\sigma}^x)^{1-j} (-\hat{\sigma}^z)^{1-k},
	\label{eq:q_tensor_unitary}
\end{align}
where Eq.\,\eqref{eq:unitary_rotations} is used. Now, we consider the $C_6$ spatial rotation and $\sigma$ reflection transformations defined in Fig.\,(1) in the main text. Such lattice symmetry transformations permute the virtual indices: $C_6 \circ (ijk) = (kij)$ and $\sigma \circ (ijk) = (jik)$. Let us apply those transformations on Eq.\,\eqref{eq:q_tensor_unitary},

\begin{align}
	&C_6 (\hat{U}_{C_6} \hat{Q}_{ijk} \hat{U}_{C_6}^{\dagger}) C_6^{-1} 
	= C_6 \left[ \tau_{i j k} (\hat{\sigma}^z)^{1-i} (\hat{\sigma}^x)^{1-j} (\hat{\sigma}^y)^{1-k}\right] C_6^{-1} 
	= \tau_{i j k} (\hat{\sigma}^x)^{1-i} (\hat{\sigma}^y)^{1-j} (\hat{\sigma}^z)^{1-k} = \hat{Q}_{ijk}, \nonumber\\
	&\sigma(\hat{U}_{\sigma} \hat{Q}_{ijk} \hat{U}_{\sigma}^{\dagger}) \sigma^{-1}
	= \sigma \left[ \tau_{i j k } (-\hat{\sigma}^y)^{1-i} (-\hat{\sigma}^x)^{1-j} (-\hat{\sigma}^z)^{1-k}\right] \sigma^{-1}
	= \tau_{i j k} (-\hat{\sigma}^x)^{1-i} (-\hat{\sigma}^y)^{1-j} (-\hat{\sigma}^z)^{1-k},
	\label{eq:q_operator_symmetric}
\end{align}
where we use the fact that the $\tau$-tensor is fully symmetric under any permutation. The first equation in Eq.\,\eqref{eq:q_operator_symmetric} directly indicates that the operator $\hat{Q}_{ijk}$, and thus the LG operator, is invariant under the $(C_6 \hat{U}_{C_6})$-transformation, $(C_6 \hat{U}_{C_6}) \hat{Q}_{\rm LG}(C_6 \hat{U}_{C_6})^{-1} = \hat{Q}_{\rm LG}$. Note that the length of any loop on the honeycomb lattice is even, which indicates that the extra minus signs in the second equation in Eq.\,\eqref{eq:q_operator_symmetric} are redundant. Therefore, the LG operator remains invariant under the $(\sigma \hat{U}_{\sigma})$-transformation: $(\sigma \hat{U}_{\sigma}) \hat{Q}_{\rm LG} (\sigma \hat{U}_{\sigma})^{-1} = \hat{Q}_{\rm LG}$. Now, we consider the time-reversal transformation $\mathcal{T}$ which transforms the operator $\hat{Q}_{ijk}$ as follows:
\begin{align}
	\mathcal{T} \hat{Q}_{ijk} \mathcal{T} 
		= \begin{cases}
		\hat{Q}_{ijk} & {\rm if }\quad i+j+k=0\\
		- \hat{Q}_{ijk}  & {\rm if }\quad  i+j+k = 2
	\end{cases},
	\label{eq:q_operator_time}
\end{align}
where  $\mathcal{T} \hat{\sigma}^\gamma \mathcal{T} = -\hat{\sigma}^\gamma$ is used. Even though the $Q$-tensor is not symmetric under the time-reversal transformation, an additional gauge transformation can restore its original form, i.e.,

\begin{align}
	\mathcal{T} \hat{Q}_{ijk} \mathcal{T} = W_{ii'} W_{jj'} W_{kk'}\hat{Q}_{i'j'k'}
	\quad {\rm with} \quad
	W = \begin{pmatrix}
		1 & 0\\ 0 & i
	\end{pmatrix}.
\end{align}
Therefore, the LG operator is time-reversal symmetric: $\mathcal{T} \hat{Q}_{\rm LG} \mathcal{T} = \hat{Q}_{\rm LG}$. By construction, the translational symmetry is respected in the LG operator, and therefore it keeps all symmetries of the isotropic Kitaev honeycomb model.

\section{Details on the Dimer gas operator}
%

%
%

As demonstrated in the main text, in order to reduce the energy, we define the DG operator $\hat{R}_{\rm DG} = {\rm tTr} \prod_{\alpha} \hat{R}_{i_\alpha j_\alpha k_\alpha} $ with
\begin{align}
	\hat{R}_{ijk} = \zeta_{i j k}(\hat{\sigma}^x)^i (\hat{\sigma}^y)^j (\hat{\sigma}^z)^k,
	\label{eq:r_operator}
\end{align}
and $i,j,k=0,1$, and the non-zero elements of $\zeta$-tensor are
\begin{align}
	\zeta_{000}=1,\quad \zeta_{100}=\zeta_{010}=\zeta_{001}=c.
\end{align}
The constant $c$ is a variational parameter. As shown in the previous section, let us first consider the $C_6 \hat{U}_{C_6}$-symmetry, i.e., $\hat{R}_{ijk} \rightarrow (C_6 \hat{U}_{C_6}) \hat{R}_{ijk} (C_6 \hat{U}_{C_6})^{-1}$:
\begin{align}
	(C_6 \hat{U}_{C_6}) \hat{R}_{ijk} (\hat{U}_{C_6}^{\dagger} C_6^{-1})
	= \zeta_{i j k} (C_6 \hat{U}_{C_6}) \left[ (\hat{\sigma}_x)^i (\hat{\sigma}_y)^j (\hat{\sigma}_z)^k \right] (\hat{U}_{C_6}^{\dagger} C_6^{-1})
	= \zeta_{i j k} C_6 \left[ (\hat{\sigma}_z)^i (\hat{\sigma}_x)^j (\hat{\sigma}_y)^k \right] C_6^{-1} = \hat{R}_{ijk},
\end{align}
where we used the fact that the $\zeta$-tensor is invariant under any permutation in the indices. Above relation implies $(C_6 \hat{U}_{C_6}) \hat{R}_{\rm DG} (C_6 \hat{U}_{C_6})^{-1} = \hat{R}_{\rm DG}$. Next, under the $(\sigma\hat{U}_\sigma)$-symmetry, the $\hat{R}_{ijk}$ transforms as

\begin{align}
	(\sigma \hat{U}_{\sigma}) \hat{R}_{ijk} (\hat{U}_{\sigma}^{\dagger} \sigma^{-1})
	&= \zeta_{i j k} (\sigma \hat{U}_{\sigma}) \left[ (\hat{\sigma}_x)^i (\hat{\sigma}_y)^j (\hat{\sigma}_z)^k \right] (\hat{U}_{\sigma}^{\dagger} \sigma^{-1})
	= \zeta_{i j k} \sigma \left[ (-\hat{\sigma}_y)^i (-\hat{\sigma}_x)^j (-\hat{\sigma}_z)^k \right] \sigma^{-1} \nonumber\\
	& = \zeta_{i j k} (-\hat{\sigma}_x)^i (-\hat{\sigma}_y)^j (-\hat{\sigma}_z)^k.
\end{align}
Here, although an extra minus sign appears, it will be canceled after after contraction since the dimer is a two-site object. Consequently, the DG operator is symmetric under the $(\sigma \hat{U}_\sigma)$-transformation: $(\sigma \hat{U}_\sigma) \hat{R}_{\rm DG} (\sigma \hat{U}_\sigma)^{-1} = \hat{R}_{\rm DG}$. As for the time-reversal symmetry, one can show

\begin{align}
	\mathcal{T} \hat{R}_{ijk} \mathcal{T} = \begin{cases}
		W_{ii'} W_{jj'} W_{kk'} \hat{R}_{i'j'k'} & {\rm if} \quad $c$ {\rm\,\,is\,\,real} \\
		\hat{R}_{ijk} & {\rm if} \quad $c$ {\rm\,\,is\,\,imaginary} 
	\end{cases}
\end{align}
Therefore, in any case, the DG operator is invariant under the time-reversal transformation: $\mathcal{T} \hat{R}_{\rm DG} \mathcal{T} = \hat{R}_{\rm DG}$. Consequently, if one can apply the PEPO on some ansatz respecting the symmetries, then the resulting state is also guaranteed to satisfy those symmetries.

%

%
\section{ The time-reversal symmetry and $(\sigma \hat{U}_{\sigma})$-symmetry of ansatze}
\begin{figure*}[!b]
  \includegraphics[width=1\textwidth]{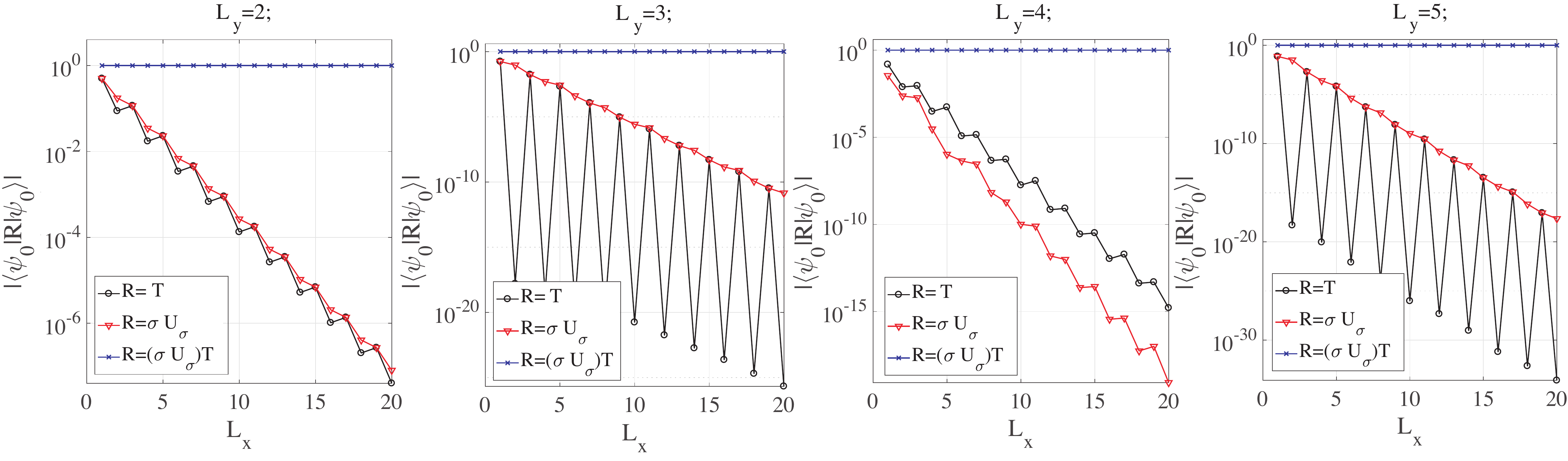}
  \caption{ Fidelity between the $|\psi_0\rangle$ and its symmetry transformed one $R|\psi_0\rangle$ on a torus with size $(L_x, L_y)$. }
  \label{fig:d2_symmetry_overlap}
\end{figure*}

The zeroth order ansatz $|\psi_0\rangle$ is obtained by contracting the zeroth order tensor
\begin{align}
	|T^0_{ijk}\rangle = \hat{Q}_{ijk} |(111)\rangle.
	\label{eq:parent_tensor}
\end{align}
Let us see how it transforms under the $(\sigma \hat{U}_{\sigma})$-symmetry and time-reversal symmetry $\mathcal{T}$,
\begin{align}
	\sigma \hat{U}_{\sigma} |T^0_{ijk}\rangle 
	= e^{i\frac{\pi}{4}}\tau_{ijk} (-\hat{\sigma}^x)^{1-i} (-\hat{\sigma}^y)^{1-j} (-\hat{\sigma}^z)^{1-k} |(-1,-1,-1)\rangle, \nonumber\\
	\mathcal{T} |T^0_{ijk}\rangle 
	= e^{i\pi} W_{ii'}W_{jj'}W_{kk'}\hat{Q}_{i'j'k'} |(-1,-1,-1)\rangle,
	\label{eq:tr_symmetry}
\end{align}
where $|(-1,-1,-1)\rangle$ denotes a spin aligned along $(-1,-1,-1)$ direction: $\langle(-1,-1,-1)|\,\vec{\sigma}\,|(-1,-1,-1)\rangle = (-1,-1,-1)/\sqrt{3}$. Here, relations $\mathcal{T}|(111)\rangle = e^{i\pi}|(-1,-1,-1)\rangle$, $\hat{U}_\sigma |(111)\rangle = e^{i\pi/4}|(-1,-1,-1)\rangle$ and $(\mathcal{T})^2 = -1$ are used. Note that one cannot restore $\mathcal{T}	|T^0_{ijk}\rangle$ and  $\sigma\hat{U}_{\sigma}|T^0_{ijk}\rangle$ to $|T^0_{ijk}\rangle$ by applying a gauge transformation. Consequently, the zeroth order tensor does not ensure the resulting state $|\psi_0\rangle$ to be time-reversal symmetric and $(\sigma \hat{U}_{\sigma})$-symmetric. However, it is invariant under the combination of $(\sigma\hat{U}_{\sigma})$ and $\mathcal{T}$ transformations, i.e., $(\sigma \hat{U}_\sigma)\,\mathcal{T} |\psi^0\rangle = e^{i\theta}|\psi^0\rangle $. More precisely, the zeroth order tensor is transformed as follows 

\begin{align}
	(\sigma \hat{U}_\sigma) \mathcal{T} |T_{ijk}^0\rangle 
	= - (\sigma \hat{U}_\sigma) \mathcal{T} \hat{Q}_{ijk} \mathcal{T} (\sigma \hat{U}_\sigma)^{-1} (\sigma \hat{U}_\sigma) \mathcal{T} |(111)\rangle
	= e^{-i\frac{\pi}{4}}\tau_{i j k} (-\hat{\sigma}^x)^{1-i} (-\hat{\sigma}^y)^{1-j} (-\hat{\sigma}^z)^{1-k} |(111)\rangle,
\end{align}
where Eqs.\,\eqref{eq:q_operator_symmetric} and \eqref{eq:q_operator_time} are used. The overall phase does not affect the resulting state, and therefore the zeroth order ansatz is invariant under the $(\sigma\hat{U}_{\sigma})\mathcal{T}$-transformation.

Even though the building block tensor does not guarantee the $\mathcal{T}$ and $\sigma \hat{U}_\sigma$ symmetries, those symmetries might be restored in a larger unit-cell. In order to carve this out, we measure the fidelity between the states $|\psi_0\rangle$ and its transformed one $O|\psi_0\rangle$ on a torus geometry with size $(L_x, L_y)$, where $O = \mathcal{T}, \sigma \hat{U}_\sigma$ and $(\sigma \hat{U}_\sigma) \mathcal{T}$, and $(L_x, L_y)$ denotes the number of unit-cell on horizontal and vertical directions. The results are shown in Fig.\,\ref{fig:d2_symmetry_overlap}. As one can see, the transformed state becomes orthogonal to each other, and thus not symmetric under the transformations. Then, how can we make them symmetric?
In fact, Eq.\,\eqref{eq:tr_symmetry} provides a quick cure to construct an ansatz respecting both the $\mathcal{T}$ and $\sigma \hat{U}_\sigma$ symmetries by doubling the bond dimension:
%
\begin{align}
	|\widetilde{T}_{ijk}^0\rangle = \eta_{i_1 j_1 k_1}^0 |T_{i_0 j_0 k_0}^0\rangle + \eta_{i_1 j_1 k_1}^1 |\mathcal{T}T_{i_0 j_0 k_0}^0\rangle,
\end{align}
where $i = 2i_1 + i_0$ and 
\begin{align}
	\eta_{i j k}^z = \begin{cases}
		1 \quad &{\rm if}\quad i+j+k=z\\
		0 \quad &{\rm others}
	\end{cases}.
	\label{eq:eta}
\end{align}
The resulting state from above tensor is simply $|\widetilde{\psi}_0\rangle = |\psi_0\rangle + |\mathcal{T} \psi_0\rangle$ and thus, by construction, invariant under the time-reversal operation. The $\sigma \hat{U}_\sigma$ symmetry is also guaranteed as shown below
%
\begin{align}
	\sigma \hat{U}_\sigma	|\widetilde{T}_{ijk}^0\rangle & = \eta_{i_1 j_1 k_1}^0 \sigma \hat{U}_\sigma |T_{i_0 j_0 k_0}^0\rangle + \eta_{i_1 j_1 k_1}^1 \sigma \hat{U}_\sigma | \mathcal{T} T_{i_0 j_0 k_0}^0\rangle\nn
	& = e^{-i\frac{\pi}{4}} \left( \eta_{i_1 j_1 k_1}^0  |\mathcal{T}T_{i_0 j_0 k_0}^0\rangle - \eta_{i_1 j_1 k_1}^1 | T_{i_0 j_0 k_0}^0\rangle \right),
\end{align}
which is restored to its original form by flipping $0 \leftrightarrow 1$ in $i_1,j_1,k_1$ with a proper coefficient to eliminate the minus sign in the second line. It can be simply done by applying a gauge trasformaion
\begin{align}
	g_\sigma = i\hat{\sigma}^y \otimes \mathbb{I}_2,
\end{align}
such that
\begin{align}
	(g_\sigma)_{ii'} (g_\sigma)_{jj'} (g_\sigma)_{kk'} \sigma \hat{U}_\sigma	 |\widetilde{T}_{i'j'k'}^0\rangle = e^{-i\frac{\pi}{4}}|\widetilde{T}_{i'j'k'}^0\rangle.
\end{align}
This procedure also applies to the higher order tensors, and therefore one can always construct a higher order ansatz, which is time-reversal and $(\sigma \hat{U}_{\sigma})$-symmetric, by doubling the bond dimension. 


%
\section{ Manipulation of $Z_2$ vortices }
\begin{figure}[!t]
  \includegraphics[width=0.6\textwidth]{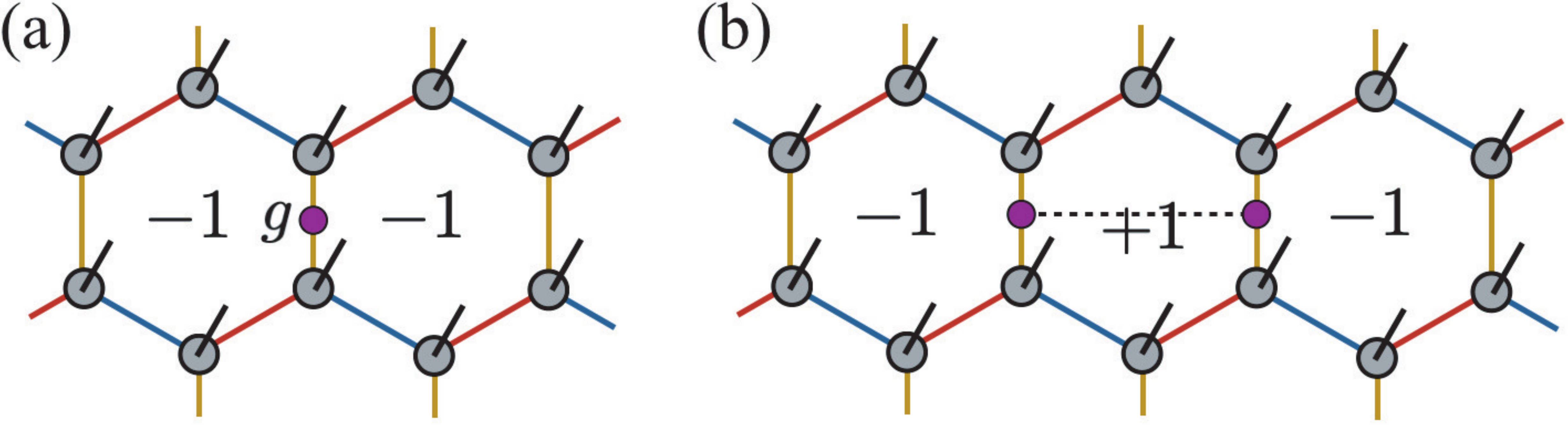}
  \caption{ (a) Insertion of non-trivial element of IGG $g$ on a bond creates two $Z_2$ vortices. (b) Additional insertion moves the $Z_2$ vortex. }
  \label{fig:flux_creation}
\end{figure}

As mentioned in the main text, one can create the $Z_2$ vortices by acting the non-trivial elements of $Z_2$ invariant guage group\,(IGG) $g$ on the virtual bonds. Let us see how it happens. For simplicity, we only consider the zeroth order tensor with $g_0=\hat{\sigma}^z$, but its generalization to the general case is straightforward. In the main text, we explicitly showed that a plaquette patch of the LG operator is invariant under the action of flux operator using Eq.\,(3) in the main text. Let us consider the same tensor network except that the $g_0$ is inserted between two sites as depicted below:  
\begin{align}
	\includegraphics[width=0.55\textwidth]{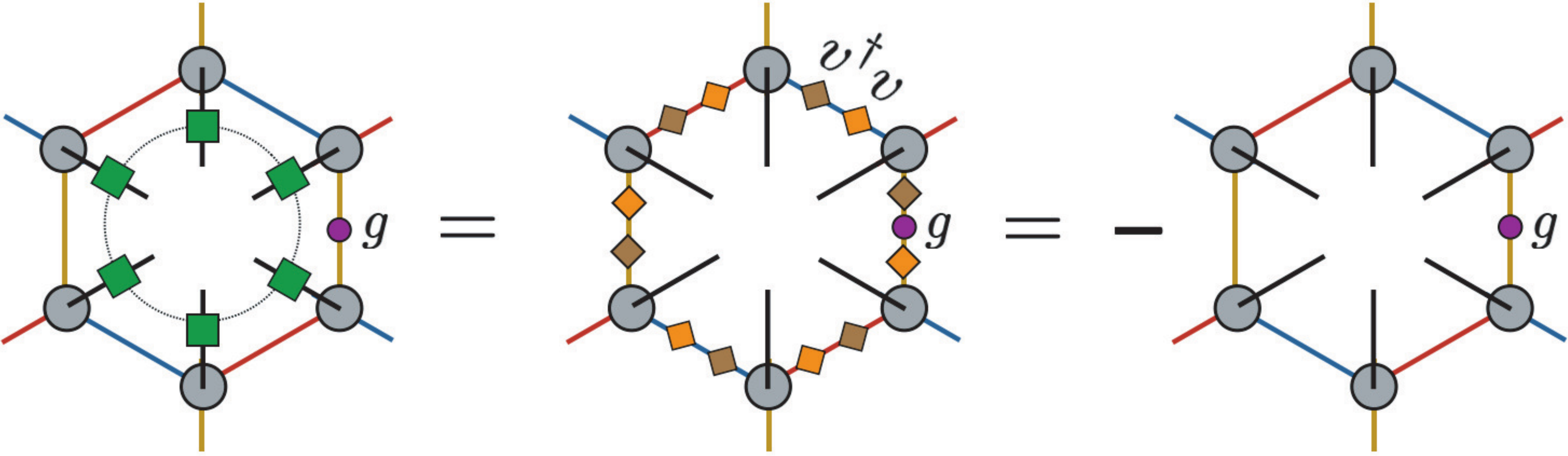},\nonumber
\end{align}
where the connected green squares denote the flux operator while
\begin{align}
	v = \begin{pmatrix}
		0 & i \\ 1 & 0 
	\end{pmatrix}.
\end{align}
In the last equality, we use $v^{\dagger} v = 1$ and $v^{\dagger} \hat{\sigma}^z v = - \hat{\sigma}^z$. Therefore, the resulting state is an eigenstate of the flux operator with the eigenvalue $-1$. The $Z_2$ vortex is created. Similarly, another vortex is created on the opposite plaquette covering the $g_0$-inserted bond as depicted in Fig.\,\ref{fig:flux_creation}\,(a). Inserting another $g_0$ in the plaquette, the minus sign will be canceled such that the vortex is removed. But, on the opposite plaqutte of newly $g_0$-inserted bond, another $Z_2$ vortex is created. In other words, the vortex can be moved from a plaquette to another one as demonstrated in Fig.\,\ref{fig:flux_creation}\,(b). The $\zeta$-tensor embedded in the DG operator is invariant only under the trivial gauge transformation $e$, and thus the IGG of the DG operator is the trivial IGG. It indicates that the multiplication of the DG operator does not enlarge the IGG. Therefore, one can create and move the vortices with $g_n = I_{2^n} \otimes \hat{\sigma}^z$ with the higher order tensors and ansatze.

\section{Loop Configurations}
\subsection{Deformation of loops}

\begin{figure}[!h]
  \includegraphics[width=0.65\textwidth]{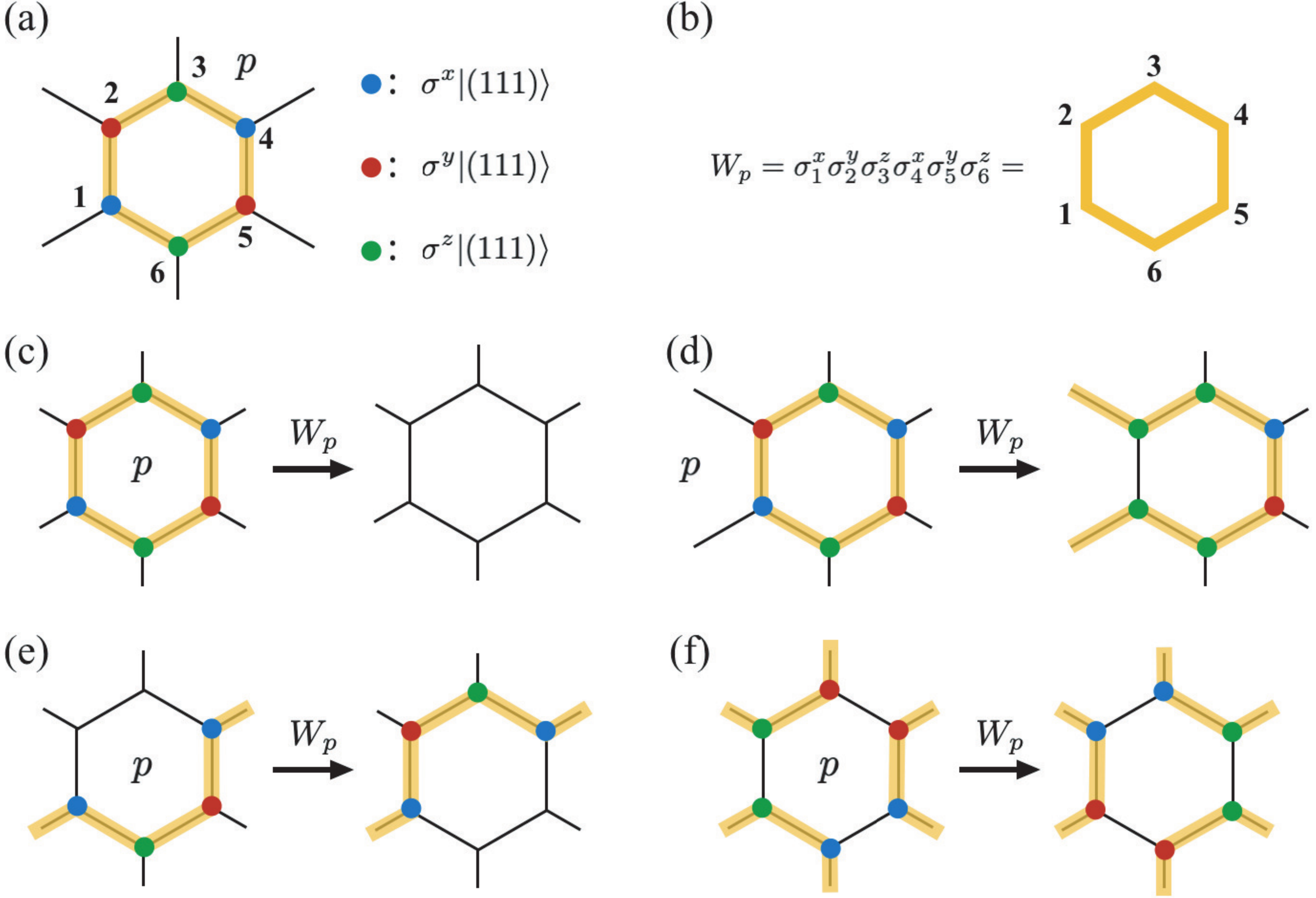}
  \caption{ Schematic figures of (a) the loop configuration on which the local state depending on the direction of the loop on each site, (b) the flux operator, and (c)-(f) the exemplary deformations of loop configurations by applying the flux operator.   }
  \label{fig:string_deformation}
\end{figure}

As mentioned in the main text, the zeroth ansatz $|\psi_0\rangle$ has the quantum loop gas structure. In other words, the ansatz are represented by linear superpositions of all possible closed loop configurations with an equal weight. Here, the loop denotes the product of $\hat{\sigma}^x|(111)\rangle, \hat{\sigma}^y|(111)\rangle$ and $\hat{\sigma}^z|(111)\rangle$ states along the loop as depicted in Fig.\,\ref{fig:string_deformation}\,(a), while an empty state is simply $|(111)\rangle$. By applying the flux operator $W_p$, one can deform the loop configurations following a simple rule. First, one regards the flux operator $W_p$ on $p$ as a loop along the boundary of $p$ as demonstrated in Fig.\,\ref{fig:string_deformation}\,(b). Then, we put $W_p$ on a loop gas configuration and draw the loop along the plaquette. If some part of loops are overlapped, then we eliminate the overlapped fragments. This procedure does not break any loop but just deform or detour the loops. Also, it does not give any extra phase after the deformation. For example, applying the flux operator on $p$ in Fig.\,\ref{fig:string_deformation}\,(a), then the $\hat{\sigma}^x$ and $\hat{\sigma}^z$ apply on $3$ and $4$, respectively. Therefore, the local states on the sites $3$ and $4$ rotates as follows:

\begin{align}
	\hat{\sigma}^z|(111)\rangle_3 \longrightarrow \hat{\sigma}^x \hat{\sigma}^z|(111)\rangle_3 = -i \hat{\sigma}^y|(111)\rangle_3, \quad 	
	\hat{\sigma}^x|(111)\rangle_4 \longrightarrow \hat{\sigma}^z \hat{\sigma}^x|(111)\rangle_4 = i \hat{\sigma}^y|(111)\rangle_4. 
\end{align}
The phases are canceled each other, and a part of loop on the bond between the sites 3 and 4 is deleted and extended to wrap the plaquette $p$. In Fig.\,\ref{fig:string_deformation}\,(c)-(f), we present some exemplary deformations of some loops by applying the flux operator. Using these local deformations, one can completely remove some loop configurations by applying the flux operators or create them from empty configurations. 

\subsection{Norm of $|\psi_0\rangle$}

The norm of wavefunction contains important informations on the low-lying excitations in the system. In this subsection, we compute the norm of $|\psi_0\rangle$ state. It is easy to see that an inner product between the loop free and an arbitrary loop configurations is simply

\begin{eqnarray}
	\langle 0 | G \rangle = \left( \frac{1}{\sqrt{3}} \right)^{l_G},
	\label{eq:inner}
\end{eqnarray}
where $|G\rangle$ and $|0\rangle = \prod |(111)\rangle$ denote respectively the arbitrary loop configuration and loop free configuration and $l_G$ is the total length of loops in the configuration $G$. Here, we used $\langle (111)| \sigma^\gamma | (111)\rangle  = 1/\sqrt{3}$. As explained in the previous subsection, any loop configuration can be obtained by a product of flux operators, i.e.,

\begin{eqnarray}
	|G\rangle = \bigotimes_{p\in G} W_p |0\rangle,
\end{eqnarray}
where $p \in G$ denotes a proper choice of plaquettes to create the $|G\rangle$ from the loop free configuration. Then, the norm of $|\psi_0\rangle$ is rewritten as

\begin{eqnarray}
	\langle \psi_0 |\psi_0\rangle = \sum_{G,G'} \langle G|G'\rangle 
	= \sum_{G,G'} \langle 0| \bigotimes_{p\in G} W_p \bigotimes_{p'\in G'} W_{p'} |0\rangle 
	= \sum_{\tilde{G}'}\sum_{\tilde{G}} \langle 0| \bigotimes_{p\in \tilde{G}} W_p |0\rangle 
	= N \sum_{G} \langle 0| G \rangle  = N \times Z_{\rm LG} \left(1,\frac{1}{\sqrt{3}}\right).
\end{eqnarray}
We use the fact that $(W_p)^2 = 1$ in the third equality, and Eq.\,\eqref{eq:inner} is substituted in the last equality. The norm of $|\psi_0\rangle$ turns out to be the partition function of the $O(n)$ loop gas model at a critical point as presented in the main text.

\subsection{Norm of $|\psi_1\rangle$}
\begin{figure}[!t]
  \includegraphics[width=1.\textwidth]{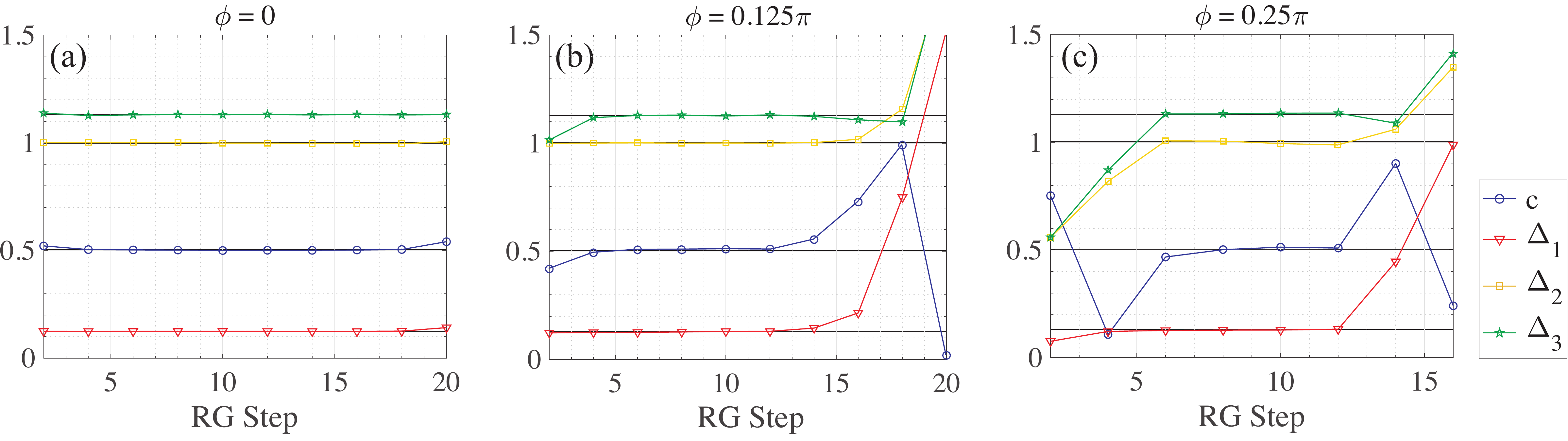}
  \caption{ Conformal data, i.e., the central charge $c$ and three largest scaling dimensions $\Delta_i$, of $\langle \psi_1(\phi) | \psi_1(\phi)\rangle$ with (a) $\phi = 0$, (b) $\phi = 0.125\pi$ and (c) $\phi = 0.25\pi$. Here, we employ the loop optimization of the tensor network renormalization with the bond dimension $\chi=48$. }
  \label{fig:rg_steps}
\end{figure}

The norm of $|\psi_1(\phi)\rangle $ does not map to the exactly solvable point of the $O(n)$ loop gas model. Therefore, we employ the loop tensor network renormalization\,(LTNR) to numerically obtain the norm of $|\psi_1(\phi)\rangle$ and extract the central charge $c$ and scaling dimensions $\Delta_i$. Results for $\phi=0,0.125\pi$ and $0.25\pi$ are shown in Fig.\,\ref{fig:rg_steps} as a function of the real space renormalization step\,(RG step). Here, the bond dimension of LTNR is fixed to $\chi=32$. The number of iteration for loop optimization varies up to 20 to find the best ansatz at each RG step. As shown in Fig.\,\ref{fig:rg_steps}\,(a), at $\phi=0$ where the tensor becomes zeroth order one, the conformal data match nicely with the exact values from Ising universality class and shows a stable behavior up to about 20 RG step. Although the accuracy of LTNR becomes less as increasing the fugacity of dimer operator\,$(\phi>0)$, we could obtain reasonable and consistent results up to around $\phi=0.25\pi$ as presented in Fig.\,\ref{fig:rg_steps}\,(b) and (c).

%

%
%
%
%

%
\section{General dimer fugacity and energy land scapes}
\begin{figure}[!b]
  \includegraphics[width=0.7\textwidth]{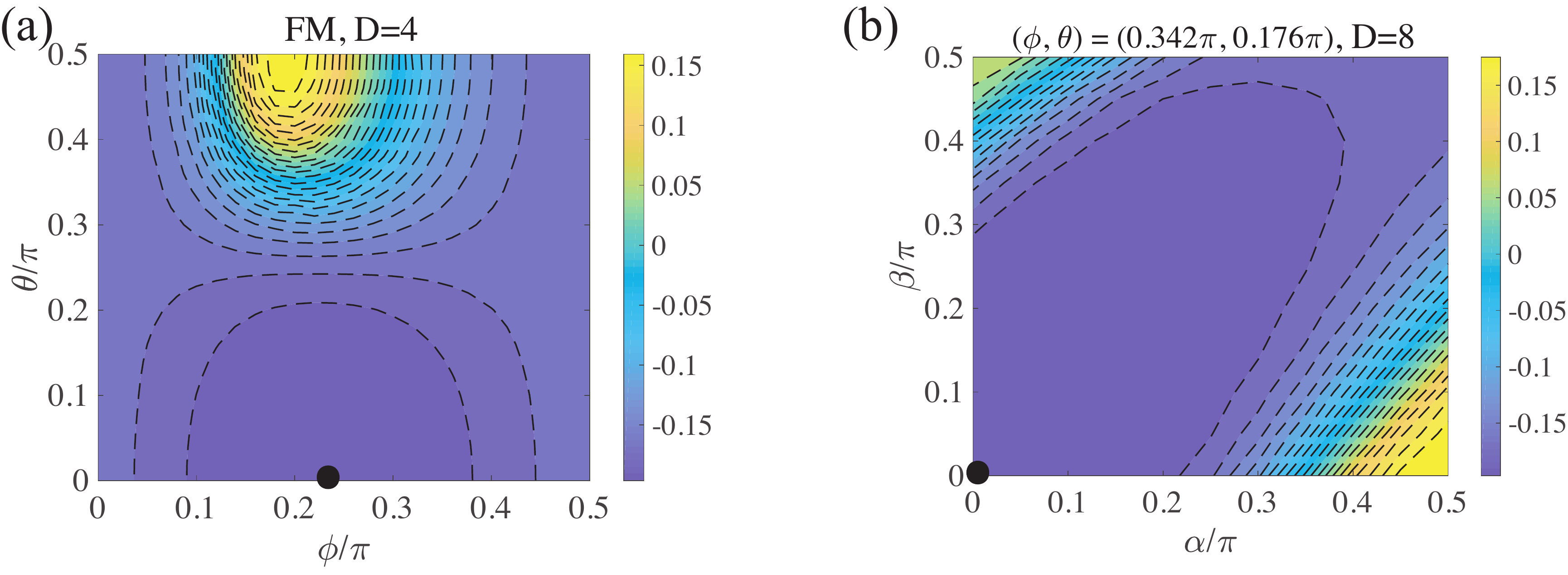}
  \caption{ Energy landscapes of ansatz defined in (a) Eq.\,\eqref{eq:complex_d4} and (b) Eq.\,\eqref{eq:complex_d8}. Black dots denote the lowest energy points, respectively. }
  \label{fig:complex_coeff}
\end{figure}

Generally, the variational parameter in the DG operator is allowed to be complex though it breaks the time-reversal symmetry. The results with only real parameters are shown in the main text. Here, the energy dependence on the complex coefficients is presented and briefly discussed. The first order ansatz $|\psi_1\rangle$ is in general defined as 
\begin{align}
	|\psi_1(\phi,\theta)\rangle = \hat{R}_{\rm DG}(\phi,\theta)|\psi_0\rangle,
	\label{eq:complex_d4}
\end{align}
where the $\zeta$-tensor in the tensor $\hat{R}_{ijk}$ is parameterized as follows: $\zeta_{000} = \cos\phi$ and $\zeta_{100}=\zeta_{010}=\zeta_{001}=e^{i\theta}\sin\phi$. Here, the parameter $\theta$ is additionally introduced to give an arbitrary phase. Now, we should fix two parameters to find the energy minimum point, and the energy landscape is shown in Fig.\,\ref{fig:complex_coeff}\,(a). The lowest energy is obtained at $(\phi,\theta) = (0.24\pi,0)$. Though we could not mathematically prove whether or not the lowest energy is found by real coefficients, we observe that real coefficients give the lowest energy at several parameter points. Now, let us consider the second order ansatz
\begin{align}
	|\psi_2(\phi,\alpha, \theta, \beta)\rangle = \hat{R}_{\rm DG}(\phi,\alpha) \hat{R}_{\rm DG}(\theta,\beta)|\psi_0\rangle.
	\label{eq:complex_d8}
\end{align}
Two phase variables $\alpha$ and $\beta$ are introduced. Therefore, we have to fix four independent parameters. Therefore, we present, in Fig.\,\ref{fig:complex_coeff}\,(b), the $\alpha$ and $\beta$ dependence of energy only at $(\phi,\theta)=(0.342\pi, 0.176\pi)$ at which the lowest energy is measured.

\section{Antiferromagnetic Kitaev Honeycomb Model}
\begin{figure}[!h]
  \includegraphics[width=0.7\textwidth]{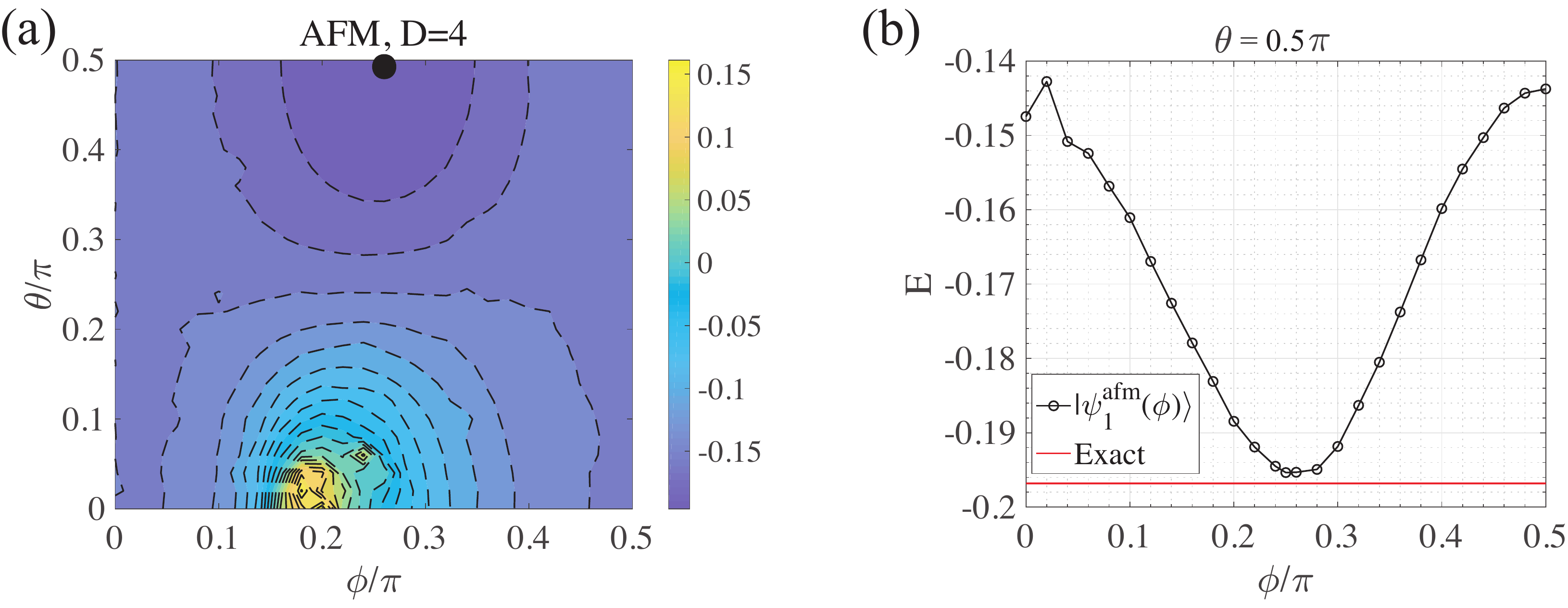}
  \caption{ Energy of (a) $|\psi_1^{\rm afm}(\phi,\theta)\rangle$ and (b) $|\psi_1^{\rm afm}(\phi,\theta=0.5\pi)\rangle$ as functions of $\phi$ and $\theta$. The lowest energy $E=-0.195356$ is found at $(\phi,\theta) = (0.25\pi,0.5\pi)$. }
  \label{fig:d4_afm_energy}
\end{figure}

In the main text, we construct the ansatze for the ferromagnetic Kitaev model by applying the LG and DG operators on the classical ground state, $|\phi^{\rm fm}\rangle = \otimes_\alpha |(111)\rangle_\alpha $ where $\alpha$ runs over all sites. Following the same strategy, we prepare the classical ground state of the antiferromagnetic model: 
\begin{align}
	|\phi^{\rm afm}\rangle = \otimes_\alpha \left[|(111)\rangle_{\alpha,a} \otimes |(-1,-1,-1)\rangle_{\alpha,b}\right],
\end{align}
where $\alpha$ labels the unit-cell, $a$ and $b$ denote two different sublattices, respectively. Then, we apply the LG and DG operators on the state $|\phi^{\rm afm}\rangle$ as we did in the ferromagnetic model and fix the variational parameters in the DG operator to find the lowest energy ansatz. The zeroth order ansatz $|\psi_0^{\rm afm}\rangle = \hat{Q}_{\rm LG} |\phi^{\rm afm}\rangle$ gives the energy $E=-0.14746$ which is rather higher than the one obtained by the zeroth ansatz of the ferromagnetic model\,(see the main text). Now, let us see how the DG operator reduces the energy. Here, we only consider the first order ansatz
\begin{align}
	|\psi_1^{\rm afm}(\phi,\theta)\rangle = \hat{R}_{\rm DG}(\phi,\theta) |\psi_0^{\rm afm}\rangle,
	\label{eq:afm_tensor}
\end{align}
where the $\zeta$-tensor in the tensor $\hat{R}_{ijk}$ is parameterized as follows: $\zeta_{000} = \cos\phi$ and $\zeta_{100}=\zeta_{010}=\zeta_{001}=e^{i\theta}\sin\phi$. The energy as functions of $\phi$ and $\theta$ is shown in Fig.\,\ref{fig:d4_afm_energy}\,(a). In contrast to the ferromagnetic case, the lowest energy is found with a negative dimer fugacity or $\theta = \pi/2$. In Fig.\,\ref{fig:d4_afm_energy}\,(b), the energy is presented as a function of $\phi$ with $\theta = 0.5\pi$. Here, the lowest energy $E=-0.195356$ is obtained at $(\phi,\theta) = (0.25\pi,0.5\pi)$, and it is only $0.75\%$ higher than the exact one. Again, this energy is slightly higher than the one obtained for the ferromagnetic model but still surprisingly close to the exact one with only $D=4$. We also confirmed that our ansatz for the antiferromagnetic model exhibit critical behavior. We believe that one could obtain much better ansatz by applying another DG operator\,($D=8$) as shown in the main text.

\section{Effect of the (111)-direction magnetic field}
\begin{figure}[!t]
  \includegraphics[width=0.7\textwidth]{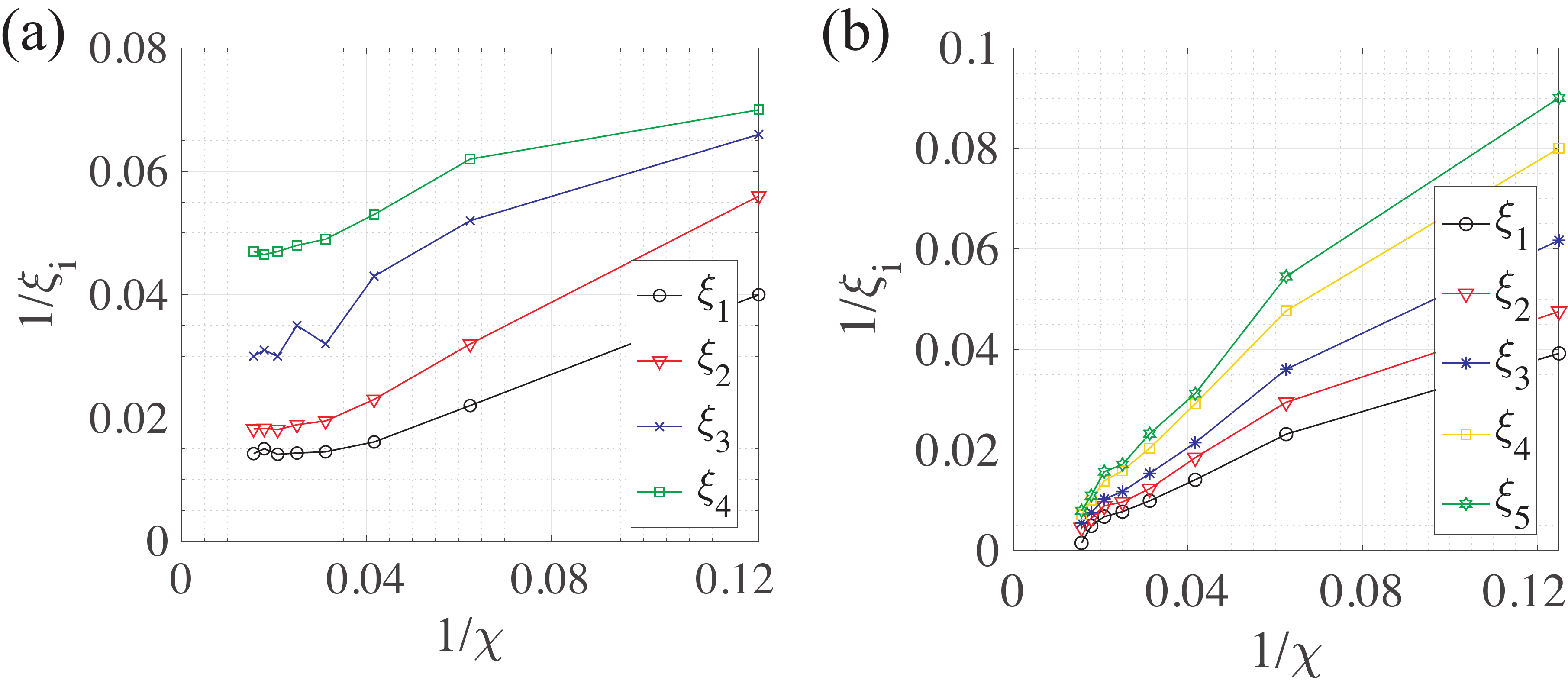}
  \caption{(a) Four longest correlation lengths in the ansatz obtained in the presence of (111)-magntic field with strength $B=0.01J$, and (b) five longest ones in the ansatz $|\psi_1(\phi=0.24\pi)\rangle$ as a function of the dimension of CTM, $\chi$. Here, the correlation length $\xi_i=\log(\lambda_0/\lambda_i)$ where $\lambda_i$ are the $(i+1)$-th largest eigenvalue of transfer matrix.}
  \label{fig:correlation_lengths}
\end{figure}

In the current representation of the KSL, it is naturally anticipated that the (111)-direction magnetic field opens the excitation gap and drives the KSL into the non-Abelian phase\cite{Kitaev2006}. Applying the field, the vortex-free condition is not required anymore, and therefore introducing a parameter $\alpha$ into the $\tau$-tensor in $\hat{Q}_{ijk}$ is allowed\cite{SM}, such that $\tau_{011} = \tau_{101} = \tau_{110} = \alpha$ and $\tau_{000}=-i$. It is obvious that $\tau_{000}$ element generates the local magnetic state $|(111)\rangle$ while the others do $\hat{\sigma}^\gamma |(111)\rangle$. In the weak-field limit, one can control $\alpha$ to modify the weight of LG and reasonably choose $\alpha<1$, since the (111)-field prefers the state $|(111)\rangle$ rather than $\hat{\sigma}^\gamma|(111)\rangle$ states. Then, the norm of wavefunction maps to $Z_{O(1)}(\alpha/\sqrt{3})$\,[Eq.\,(6) in the main text], where
the model enters into a massive phase\cite{Domb2000}.
Consequently, the gap is opened by the magnetic field. 

In order to find a better ansatz with $D=4$, we introduce two additional parameters in $|\psi_1(\phi)\rangle$, which tune the weight of local $\hat{\sigma}^\gamma |(111)\rangle$ states. To be more specific, let us explicitly write down the non-zero elements of the first order tensor $|T^1_{ijk}(\phi)\rangle = \hat{R}_{i_1 j_1 k_1}(\phi) \hat{Q}_{i_0 j_0 k_0}|(111)\rangle$ where $i=2i_1 + i_0$. For simplicity, we redefine the first order tensor as $|T^1_{ijk}(\phi)\rangle = \cos\phi|A_{ijk}\rangle + \sin\phi|B_{ijk}\rangle$ with
\begin{align}
	& |A_{000}^1\rangle = |(111)\rangle, 
	&&|A_{011}^1\rangle = \hat{\sigma}^x|(111)\rangle, 
	&&|A_{101}^1\rangle = \hat{\sigma}^y |(111)\rangle, 
	&&|A_{110}^1\rangle = \hat{\sigma}^z |(111)\rangle, \nn
	& |B_{211}^1\rangle = |(111)\rangle, 
	&&|B_{200}^1\rangle = \hat{\sigma}^x|(111)\rangle, 
	&&|B_{310}^1\rangle = -i \hat{\sigma}^y |(111)\rangle, 
	&&|B_{301}^1\rangle = i \hat{\sigma}^z |(111)\rangle, \nn
	& |B_{121}^1\rangle = |(111)\rangle, 
	&&|B_{130}^1\rangle = i \hat{\sigma}^x |(111)\rangle, 
	&&|B_{020}^1\rangle = \hat{\sigma}^y |(111)\rangle, 
	&&|B_{031}^1\rangle = -i \hat{\sigma}^z |(111)\rangle, \nn
	& |B_{112}^1\rangle = |(111)\rangle, 
	&&|B_{103}^1\rangle = -i\hat{\sigma}^x |(111)\rangle,
	&&|B_{013}^1\rangle = i \hat{\sigma}^y |(111)\rangle, 
	&&|B_{002}^1\rangle = \hat{\sigma}^z |(111)\rangle.
	\label{eq:d4_elements}
\end{align}
Note that the $A$-tensor is the same as the zeroth order tensor. Now, we assign parameters tuning the weight of $\hat{\sigma}^\gamma|(111)\rangle$ states in the $A$-tensor and $B$-tensor, respectively. In other words, the ansatz becomes dependent on three parameters: $|T_{ijk}^1(\phi,\alpha_0,\alpha_1)\rangle = \cos\phi|A_{ijk}(\alpha_0)\rangle + \sin\phi|B_{ijk}(\alpha_1)\rangle$ where the parameter $\alpha_i$ multiplied to the $\hat{\sigma}^\gamma |(111)\rangle$ states. For example, the non-zero elements of $A$-tensor are
\begin{align}
	|A_{000}^1(\alpha_0)\rangle = |(111)\rangle, 
	&&|A_{011}^1(\alpha_0)\rangle = \alpha_0\,\hat{\sigma}^x|(111)\rangle, 
	&&|A_{101}^1(\alpha_0)\rangle = \alpha_0\,\hat{\sigma}^y|(111)\rangle, 
	&&|A_{110}^1(\alpha_0)\rangle = \alpha_0\,\hat{\sigma}^z|(111)\rangle	.
\end{align}
Note that the $(C_6\hat{U}_{C_6})$-symmetry is still valid even in the presence of the (111)-field. Therefore, one allows to introduce only a single parameter $\alpha_1$ in the $B$-tensor. Since the $(111)$-field prefers the local state $|(111)\rangle$ rather than $\hat{\sigma}^\gamma|(111)\rangle$, one may naively expect that reducing the parameters $\alpha_i$ from 1 helps lowering the energy. Indeed, we found that the energy is optimized at $(\phi,\alpha_0,\alpha_1) = (0.225\pi, 0.825, 0.95)$ with the energy $E = -0.19688$ which is competitive to the one obtained by the numerical optimization\,\cite{Ryui19}.
 The norm of the ansatz is not mapped into the LG model similar to $|\psi_1(\phi)\rangle$. Therefore, in order to show its gapped nature, we directly measure the most dominant correlation lengths of the ansatz using the environment tensor in CTMRG\cite{Takahashi2005,hylee18}. The result for ansatz with $B=0.01J$ is presented in Fig.\,\ref{fig:correlation_lengths}\,(a) as a function of $\chi$. As one can see, the correlation lengths converge to finite values with increasing $\chi$ indicating a finite gap in the ansatz. For comparison, those in the critical state $|\psi_1(\phi=0.24\pi)\rangle$ are shown in Fig.\,\ref{fig:correlation_lengths}\,(b), which exhibit diverging behavior with $\chi$. Therefore, the gapped ansatz in the presence of (111)-field can be reasonably obtained by giving some fugacity to the tensor element generating $\hat{\sigma}^\gamma|(111)\rangle$ state. In addition, it has been recently shown\cite{Fendley2008} that the gapped LG having non-trivial inner products between two configurations can be systematically mapped into string-net states describing non-Abelian anyonic excitations\cite{Fendley2008}. It strongly suggests that our ansatz belongs to a non-Abelian phase in the presence of (111)-field, which is consistent with the perturbation calculation using the Majorana fermion in Kitaev's original work\,\cite{Kitaev2006}.

\section{$\chi$-scaling of the variational energies}

\begin{figure}[!t]
  \includegraphics[width=1.\textwidth]{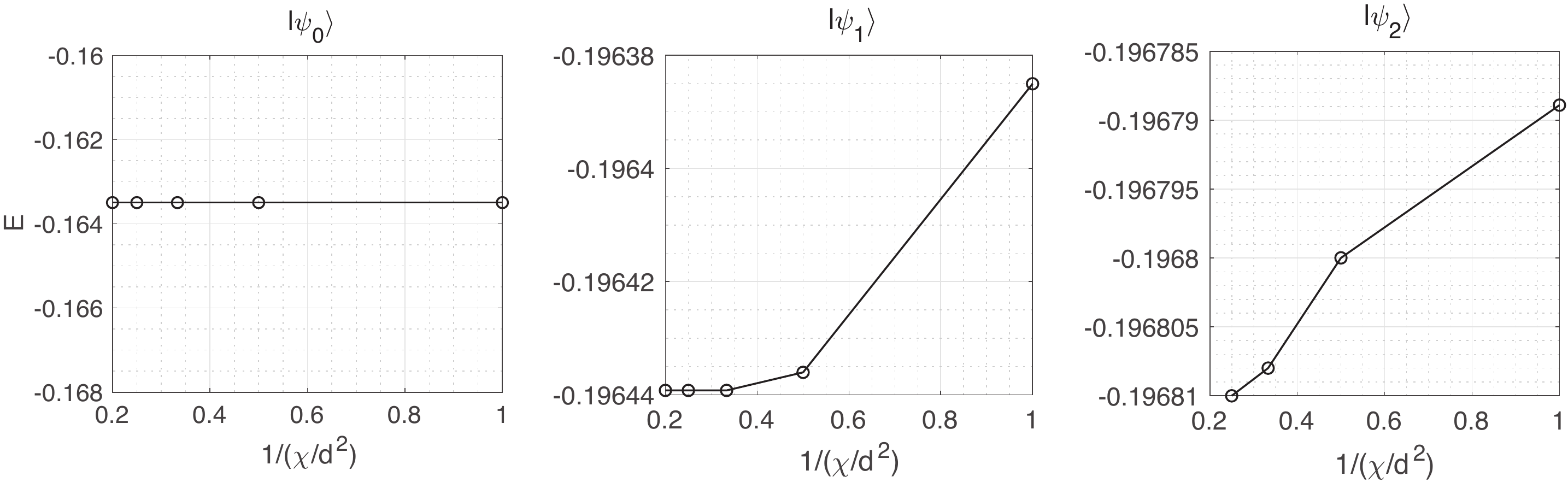}
  \caption{The $\chi$-dependence of the variational energies of the loop gas state $|\psi_0\rangle$\,($d=2$) and string gas states $|\psi_1\rangle$\,($d=4$) and $|\psi_2\rangle$\,($d=8$) from left to right, respectively. }
  \label{fig:energy_chi}
\end{figure}

We provide the bond dimension of CTMRG, $\chi$, dependence of the variational ansatze shown in the main manuscript. In Figure\,\ref{fig:energy_chi}, the scaling behavior of the variational energies of the LG state $|\psi_0\rangle$, the first order SG ansatz $|\psi_1(\phi=0.24\pi)\rangle$, and the second order SG ansatz $|\psi_2(\phi=0.342\pi, \theta=0.176\pi)\rangle$ are shown from left to right, respectively.

\end{document}